\documentclass[preprint]{aastex63}

\newcommand\COtwo{CO$_2$ }
\newcommand\SOtwo{SO$_2$ }
\newcommand{\Lagr}{\mathcal{L}}
\newcommand{\Z}{\mathcal{Z}}
\newcommand{\B}{\mathcal{B}}

\usepackage{amsmath}
\usepackage{gensymb}
\usepackage{comment}
\setcitestyle{notesep={; }}

\shorttitle{A comprehensive revisit of select Galileo/NIMS observations of Europa}
\shortauthors{Mishra et al.}

\begin{document}

\title{A comprehensive revisit of select Galileo/NIMS observations of Europa}

\correspondingauthor{Ishan Mishra}
\email{im356@cornell.edu}

\author[0000-0001-6092-7674]{Ishan Mishra}
\affiliation{Department of Astronomy and Carl Sagan Institute, Cornell University \\
122 Sciences Drive \\
Ithaca, NY 14853, USA}

\author[0000-0002-8507-1304]{Nikole Lewis}
\affiliation{Department of Astronomy and Carl Sagan Institute, Cornell University \\
122 Sciences Drive \\
Ithaca, NY 14853, USA}

\author[0000-0003-2279-4131]{Jonathan Lunine}
\affiliation{Department of Astronomy and Carl Sagan Institute, Cornell University \\
122 Sciences Drive \\
Ithaca, NY 14853, USA}

\author[0000-0002-3225-9426]{Kevin P. Hand}
\affiliation{Jet Propulsion Laboratory, California Institute of Technology, \\
Pasadena, CA 91109, USA}

\author[0000-0003-4870-1300]{Paul Helfenstein}
\affiliation{Cornell Center for Astrophysics and Planetary Science, Cornell University \\
122 Sciences Drive \\
Ithaca, NY 14853, USA}

\author{R. W. Carlson}
\affiliation{Jet Propulsion Laboratory, California Institute of Technology, \\
Pasadena, CA 91109, USA}

\author[0000-0003-4816-3469]{Ryan J. MacDonald}
\affiliation{Department of Astronomy and Carl Sagan Institute, Cornell University \\
122 Sciences Drive \\
Ithaca, NY 14853, USA}



\begin{abstract}

The \emph{Galileo} Near Infrared Mapping Spectrometer (NIMS) collected spectra  of Europa in the 0.7-5.2 $\micron$ wavelength region, which have been critical to improving our understanding of the surface composition of this moon. However, most of the work done to get constraints on abundances of species like water ice, hydrated sulfuric acid, hydrated salts and oxides have used proxy methods, such as absorption strength of spectral features or fitting a linear mixture of laboratory generated spectra. Such techniques neglect the effect of parameters degenerate with the abundances, such as the average grain-size of particles, or the porosity of the regolith. In this work we revisit three \emph{Galileo} NIMS spectra, collected from observations of the trailing hemisphere of Europa, and use a Bayesian inference framework, with the Hapke reflectance model, to reassess Europa's surface composition. Our framework has several quantitative improvements relative to prior analyses: (1) simultaneous inclusion of amorphous and crystalline water ice, sulfuric-acid-octahydrate (SAO), CO$_2$, and SO$_2$; (2) physical parameters like regolith porosity and radiation-induced band-center shift; and (3) tools to quantify confidence in the presence of each species included in the model, constrain their parameters, and explore solution degeneracies. We find that SAO strongly dominates the composition in the spectra considered in this study, while both forms of water ice are detected at varying confidence levels. We find no evidence of either CO$_2$ or SO$_2$ in any of the spectra; we further show through a theoretical analysis that it is highly unlikely that these species are detectable in any 1-2.5 $\micron$ \emph{Galileo} NIMS data.

\end{abstract}

\keywords{Jovian satellites -- Bayesian statistics -- Near Infrared Astronomy -- Europa}


\vspace{5 pt}

\section{Introduction} \label{sec:intro}

Europa's young surface and geological features like chaos terrains suggest an active exchange of materials between the surface and subsurface ocean \citep{carr_evidence_1998, moore_surface_2009, kattenhorn_evidence_2014, trumbo_sodium_2019}. Accessing the ocean directly to determine its composition will require a sustained program of scientific and technological developments \citep{hand_exploring_2018}, and thus our best near-term prospect for determining the ocean composition and constraining Europa's habitability is to understand its surface composition with available spectra. Europa's icy surface is, at present, our primary window into the chemistry of the putative subsurface ocean.  

Most of our knowledge of Europa's surface composition comes from reflectance spectroscopy, using ground- and space-based telescopes, as well as instruments onboard spacecraft. The near-infrared reflectance data collected by the \emph{Galileo} mission, specifically from the Near-Infrared Mapping Spectrometer \citep{carlson_near-infrared_1992}, has revolutionized our understanding of Europa's surface \citep[e.g.][]{mccord_salts_1998, carlson_hydrogen_1999, mccord_hydrated_1999, hansen_amorphous_2004, carlson_distribution_2005, hansen_widespread_2008, dalton_europas_2012, shirley_europas_2016}. Spectroscopy in the near-infrared has helped reveal the majority of species identified on Europa's surface \citep{carlson_europas_2009}, including hydrated sulfuric acid \citep{carlson_sulfuric_1999, carlson_distribution_2005}, hydrated sulfates \citep{mccord_salts_1998, dalton_linear_2007}, chlorinates \citep{fischer_spatially_2015, ligier_vlt/sinfoni_2016} and oxidants \citep{hansen_widespread_2008, hand_keck_2013}. Although serendipitous observations of Europa by \textit{Juno} \citep{filacchione_serendipitous_2019, mishra_bayesian_2021} have provided a new window into this world, the \emph{Galileo} NIMS dataset still holds great potential to further improve our understanding of Europa's surface composition, by the application of new, comprehensive methods, as elucidated below. 

A standard approach to finding a species in reflectance data is through a spectral fitting analysis \citep[e.g.][]{carlson_distribution_2005, clark_surface_2012}. Using a radiative-transfer model, such an analysis also constrains a subset of physical properties like abundances and average grain-sizes. For the NIMS data of Europa, much work has been done with models that fit the data with a linear combination of laboratory spectra of various species \citep[e.g.][]{dalton_linear_2007, dalton_europas_2012}. However, this is a rough approximation to the highly non-linear radiative-transfer process through a planetary regolith \citep[e.g.][]{shirley_europas_2016}. As a result, linear-mixing type analyses do not allow us to disentangle key properties of the regolith such as average grain-size, porosity, observational geometry parameters, and species abundances, which are highly degenerate with each other. 

A more physically motivated alternative is a thorough radiative-transfer model that uses fundamental optical properties of each end-member, alongside parameterizations for physical properties of the regolith \citep[e.g.][]{hapke_2012}. Such a model takes into account various inherent degeneracies in the radiative-transfer process and thus yields more conservative estimates of end-member properties, such as abundances and grain-sizes. Physically motivated radiative transfer schemes, such as the Hapke model \citep{hapke_2012}, require that optical constants (e.g. refractive indices and extinction coefficients) be measured for several species of interest in the relevant temperature ($\sim 100$ K) and wavelength regime ($\sim 1-5 \ \micron$ for most NIMS data). Such laboratory measurements can be challenging, which limits the potential Europan species we can explore using this technique. Thankfully, such optical constants exist for certain key species: amorphous and crystalline water ice at 120 K \citep{grundy_temperature-dependent_1998, mastrapa_optical_2009}, CO$_2$ at 179 K \citep{quirico_near-infrared_1997, quirico_composition_1999, sshade}, SO$_2$ at 125 K \citep{schmitt_identification_1994, schmitt_optical_1998, sshade} and sulfuric-acid-octahydrate (hereafter SAO) at 77 K \citep{carlson_distribution_2005}. 

SAO is a major product of the radiolytic sulfur cycle on the trailing hemisphere of Europa \citep{carlson_europas_2009}. \citet{carlson_distribution_2005} used NIMS data and a simple two-component reflectance model of crystalline ice and SAO to map the distribution of SAO on the leading and trailing hemispheres of Europa, finding concentrations as high as 90\% near the trailing hemisphere apex. However, an important form of water ice in the upper-most layers of Europa's surface may be amorphous ice \citep[e.g][]{hansen_widespread_2008}, which was not included in the work by \citet{carlson_distribution_2005} due to the unavailability of its optical constants at that time. Since the study of \citet{carlson_distribution_2005}, a comprehensive database of cryogenic water-ice optical constants, in both amorphous and crystalline forms, was published by \citet{mastrapa_optical_2009}. 

The trace oxides CO$_2$ and SO$_2$ are astrobiologically significant oxidants on Europa’s surface and might play an important role in creating an environment of chemical disequilibrium in its subsurface ocean \citep[e.g][]{chyba_possible_2001, hand_energy_2007}. Moreover, CO$_2$ has also been linked to young geological regions on Europa like the chaos terrains \citep{trumbo_h2o2_2019}, which indicates a possibly endogenic origin of the carbon that fuels a carbon-cycle on Europa's surface \citep{carlson_europas_2009,hand2012carbon}. Published estimates of CO$_2$ abundances exist for the leading side of Europa, where the 4.25 $\micron$ feature in NIMS spectra \citep{hand_energy_2007} was used to constrain the abundance to be $\sim$ 360 ppm. For SO$_2$, a rough abundance estimate of $\sim 0.2\%$ comes from disk-integrated UV observations of Europa \citep{hendrix_europas_2008, carlson_europas_2009} -- although a 4.0 $\micron$ feature has been identified in NIMS spectra \citep{hansen_widespread_2008}. Theoretical studies of CO$_2$ and SO$_2$ clathrate formation indicate that the bulk ice-shell of Europa could have an oxidant concentration of up to $\sim 7\%$ \citep{hand_clathrate_2006}. It should be noted that while sulfur and its allotropes are also present in significant amounts on the trailing side of Europa -- due to the Io-genic bombardement \citep{carlson_europas_2009} -- they lack strong features in the NIR wavelength region and hence can be assumed to not affect the NIMS spectra. 

To date, only water-ice and SAO have been simultaneously considered in a radiative-transfer based analysis of NIMS data. However, Europa-specific optical constants are now available for amorphous ice, crystalline ice, SAO, CO$_2$, and SO$_2$. Hence, there is an opportunity to model all five species simultaneously in a revised analysis of Galileo NIMS observations of Europa.

In this work we re-analyze a subset of NIMS spectra using the Hapke reflectance model \citep{hapke_2012} in a Bayesian inference framework. Bayesian inference has gained popularity in planetary surface spectroscopic inversion in recent years \citep[e.g.][]{fernando_surface_2013, schmidt_realistic_2015, fernando_martian_2016, lapotre_probabilistic_2017, rampe_sand_2018, belgacem_regional_2020}. The goal of this work is to demonstrate a spectroscopic analysis approach rooted in Bayesian inference \citep{mishra_bayesian_2021} that permits several advances over previous analyses of Europan data, specifically: (1) include five species (amorphous ice, crystalline ice, SAO, CO$_2$, and SO$_2$) in a radiative-transfer analysis of Europan spectra; (2) include parameters that account for physical effects, such as radiation-induced band shifts, porosity of the regolith, and calibration uncertainty of the instrument; (3) yield statistical constraints on parameters, including confidence intervals; and (4) provide detection significances for each individual species with a statistical metric via Bayesian model comparisons. The fitting analysis in our approach enables identification of trace species, taking into account their effect on the spectrum over the entire wavelength range, instead of relying on a few sharp features. 

We apply this framework to three NIMS spectra of the trailing hemisphere of Europa, spanning 1.0-2.5 $\micron$. These data were  available in their reduced form, along with all the observation geometry parameters needed for our analysis, through previous work by \citet{carlson_distribution_2005}. The data and their properties are described in section \ref{sec:data}. Section \ref{sec:methods} lays out the details of our analysis framework, including the Hapke reflectance model (section \ref{sec:forward_model}) and the Bayesian inference methodology (section \ref{sec:Bayes_sampling}). We discuss the results from the application of this framework to the three spectra in section \ref{sec:results}, followed by a theoretical study (section \ref{sec:detectability}) to gauge the detectability of CO$_2$ and SO$_2$ as a function of their abundance in synthetic NIMS spectra over the 1.0-2.5 $\micron$ range. We conclude with a discussion (section \ref{sec:discussion}) of our results, placing them in the context of existing knowledge of Europa's surface composition, before summarizing  (section \ref{sec:conclusions}) the key takeaways from our work. 

\section{Data} \label{sec:data}

\begin{figure}[htbp!]
\centering
\includegraphics[width=0.75\textwidth]{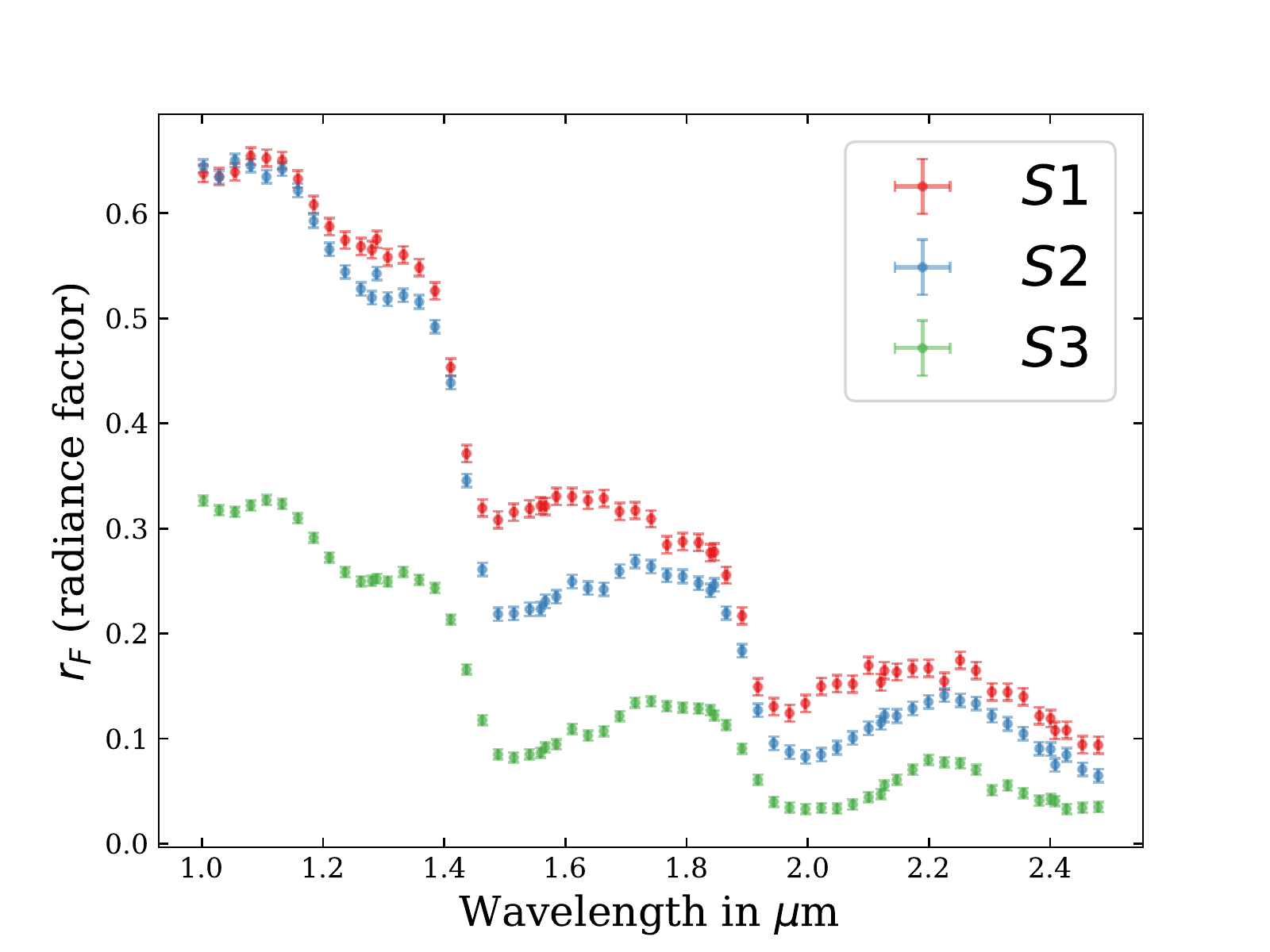}
\caption{The three Galileo NIMS observations, referred to as $S1$, $S2$ and $S3$, being used in this work. The spectra come from observation G1ENNHILAT01 by Galileo that mapped the trailing hemisphere of Europa. Details for the three spectra are provided in Table \ref{tab:data} and the their locations on Europa's surface is shown in Figure \ref{fig:europa_map}. $S1$, $S2$ and $S3$ also correspond to the bottom, middle and top spectra, respectively, shown in Figure 3 of \citet{carlson_distribution_2005}. \label{fig:data}}
\end{figure}

\begin{figure}[htbp!]
\centering
\includegraphics[width=\textwidth]{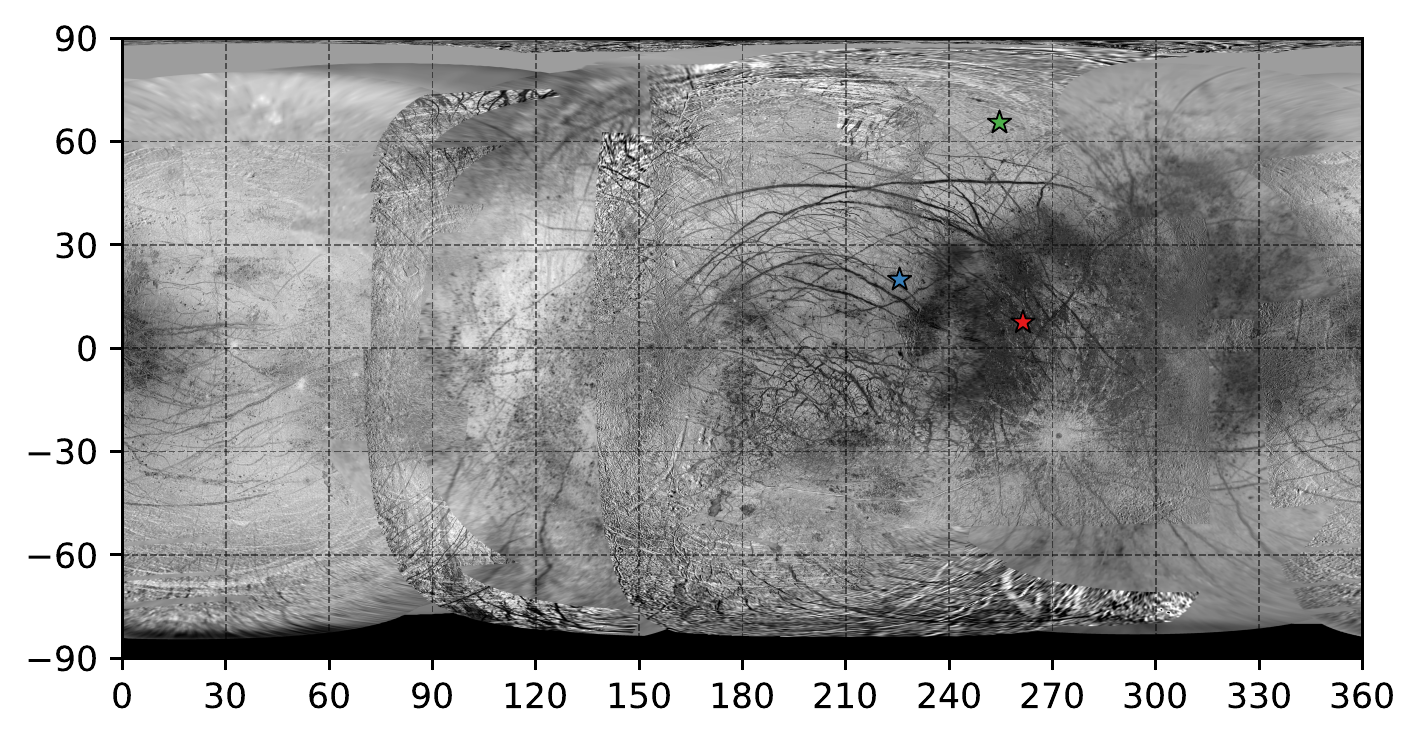}
\caption{The approximate locations on Europa's surface, marked as red, blue and green stars, corresponding to spectra $S1$ (7.5$\degr$ N, 261.4$\degr$ W), $S2$ (19.9$\degr$ N, 225.6$\degr$ W) and $S3$ (65.5$\degr$ N, 254.6$\degr$ W) respectively (see Table \ref{tab:data} for details). This map is centered on 180$\degr$ W, where the 0$\degr$ W corresponds to the sub-Jovian longitude. The global base map of Europa was downloaded from USGS `Map a Planet'$^a$, derived from Galileo and Voyager images, and is shown in cylindrical projection.}\label{fig:europa_map}
$^a${\url{https://astrocloud.wr.usgs.gov/index.php?view=map2}}
\end{figure}

During the \emph{Galileo} spacecraft's first orbit of Jupiter, the NIMS instrument acquired
several spectral maps of Europa, the first being a map of the northern hemisphere in
the trailing- side anti-Jupiter region and termed the G1ENNHILAT01A observation.
Approximately 300 spectra were obtained in this observation and projected as a
point-perspective spectral radiance-factor map. Three types of terrain (highly
hydrated, icy, and intermediate) were chosen, for illustration of the diversity in composition, and a single pixel
spectrum for each region was extracted and fit using experimentally-derived indices
of refraction \citep[][note the interchange of radiance factor and coefficient therein]{carlson_distribution_2005}. This work focuses on those three spectra, which are shown in Figure \ref{fig:data}, and will henceforth be referred to as $S1$, $S2$ and $S3$. The locations on Europa's surface corresponding to these three spectra are shown in Figure \ref{fig:europa_map}, and details about the spacecraft observation geometry is shown in Table \ref{tab:data}. \citet{carlson_distribution_2005} found that $S1$, $S2$ and $S3$ represent a range of compositions of Europa's trailing hemisphere, from highly hydrated to minimally hydrated. Hence, they also allow us to explore trends in the physical and chemical parameters included in our model across regions of differing water-ice content.

\begin{deluxetable*}{cccchcDc}[htbp!]
\tablenum{1}
\tablecaption{NIMS spectra from the G1ENNHILAT01 observation of Europa being analyzed in this work and referred to as $S1$, $S2$ and $S3$. These spectra are shown in Figure \ref{fig:data}.}
\tablewidth{0pt}
\tablehead{
\colhead{Spectrum} & \colhead{Line-sample} & \colhead{Latitude} & \colhead{Longitude} & \nocolhead{Deg. min.} & \colhead{Incidence} & \multicolumn2c{Emergence} & \colhead{Phase}\\
\colhead{} & \colhead{} & \colhead{} & \colhead{} & \nocolhead{} & \colhead{angle} & \multicolumn2c{angle} & \colhead{angle}}
\decimals
\startdata
S1 & 32-27 & 7.5 & 261.4 & 96$\degr$ 18' E, 17$\degr$ 24' N & 24.3 & 43.8 & 31.0 \\
S2 & 45-46 & 19.9 & 225.6 & 134$\degr$ 36' E, 17$\degr$ 54' N & 25.0 & 7.0 & 30.9 \\
S3 & 20-64 & 65.5 & 254.6 & 105$\degr$ 24' E, 65$\degr$ 30' N & 67.9 & 49.1 & 30.8
\enddata
\tablecomments{The G1ENNHILAT01 observation\footnote{The original data for this orbit was written to a CD and was submitted to the PDS as Volume `go\_1104' and file `g1e002ci.qub'. The PDS filename is now `g1e003' followed by `cr' or `ci', with `cr' and `ci' denoting radiance and radiance factor data respectively} approximately spanned latitudes 30$\degr$ S - 90$\degr$ N and longitudes 150$\degr$ W - 300$\degr$ W, with approximate scale of 39 km/pixel \citep{carlson_distribution_2005}. Latitude 0$\degr$ and Longitude 0$\degr$ correspond to the sub-Jovian point on Europa's surface, with longitude increasing positively westward towards the leading hemisphere. $S1$, $S2$ and $S3$ are all from the trailing hemisphere of Europa. \label{tab:data}}
\end{deluxetable*}

The compositional discriminative power of a spectral analysis method like Bayesian inference depends sensitively on the uncertainties in the data (see section 3 of \citet{mishra_bayesian_2021}). For most NIMS spectra, including $S1$, $S2$ and $S3$, accurate uncertainties could not be calculated because \emph{Galileo}’s satellite observations were limited in temporal duration by the relative spacecraft-satellite velocity and by the solar illumination incidence angles. In addition, data transmission to the Earth was severely limited, consequently repeated measurements of a particular region on Europa's surface were precluded. It is the intrinsic variation among the repeated measurements that informs the uncertainties on the data. Estimates of the uncertainties on data of the kind considered here are usually obtained from repeated measurements of a particular region of interest. Because the spacecraft's flybys when the data was collected were very short in duration, such repeated measurements could not be taken.

While rough estimates on the SNR (signal-to-noise ratio) of Galileo data in the literature put the number between 5 and 50 \citep{greeley_future_2009}, the noise characteristics vary with wavelength and depend on instrumental and external factors that change from observation to observation \citep{carlson_near-infrared_1992}. The noise in the NIMS instrument includes thermal noise (which comes from the blackbody IR emission of the interior instrument housing), detector current noise that is approximately proportional to the inverse of the solar spectral irradiance, a small amount of digitization noise, and radiation noise that produces pulses of detector current with widely ranging amplitudes. The observations we are using did not suffer excessive radiation noise, but a small but non-quantifiable amount. We're assuming we're working in the photon-noise limited regime, which allows us to measure all the sources of ``error" via the scatter they induce on the signal. We obtain estimates of this inherent scatter in each of $S1$, $S2$ and $S3$, caused by a combination of all the factors listed above, by adopting the following process: 

\begin{enumerate}
    \item Assume a SNR of 50 across all wavelength channels.
    \item Perform a fitting analysis with a full model (all five species included) and evaluate the best-fit or the maximum-likelihood solution. The details of the analysis framework are discussed in the next section (section \ref{sec:methods}). 
    \item Calculate the residuals of this fit to the data. 
    \item Calculate the standard deviation of these residuals or the SDR.
    \item Repeat the steps for SNR of 5.
\end{enumerate}

The larger of the two SDR values, from the SNR=5 and SNR=50 cases, is chosen to be a conservative estimate of the noise in a given spectrum and is assigned as the uncertainty across all channels. This analysis results in data uncertainties of 0.0081, 0.0063, and 0.0047 (in radiance factor units), which correspond to an average SNR (averaged across the wavelength channels) of 39.51, 44.43 and 28.35 for $S1$, $S2$ and $S3$ respectively. These values fall in the range of 5 to 50 that is generally assumed to be the average SNR of NIMS spectra of Europa \citep{greeley_future_2009}.
 
\begin{figure}
\centering
\includegraphics[width=\textwidth]{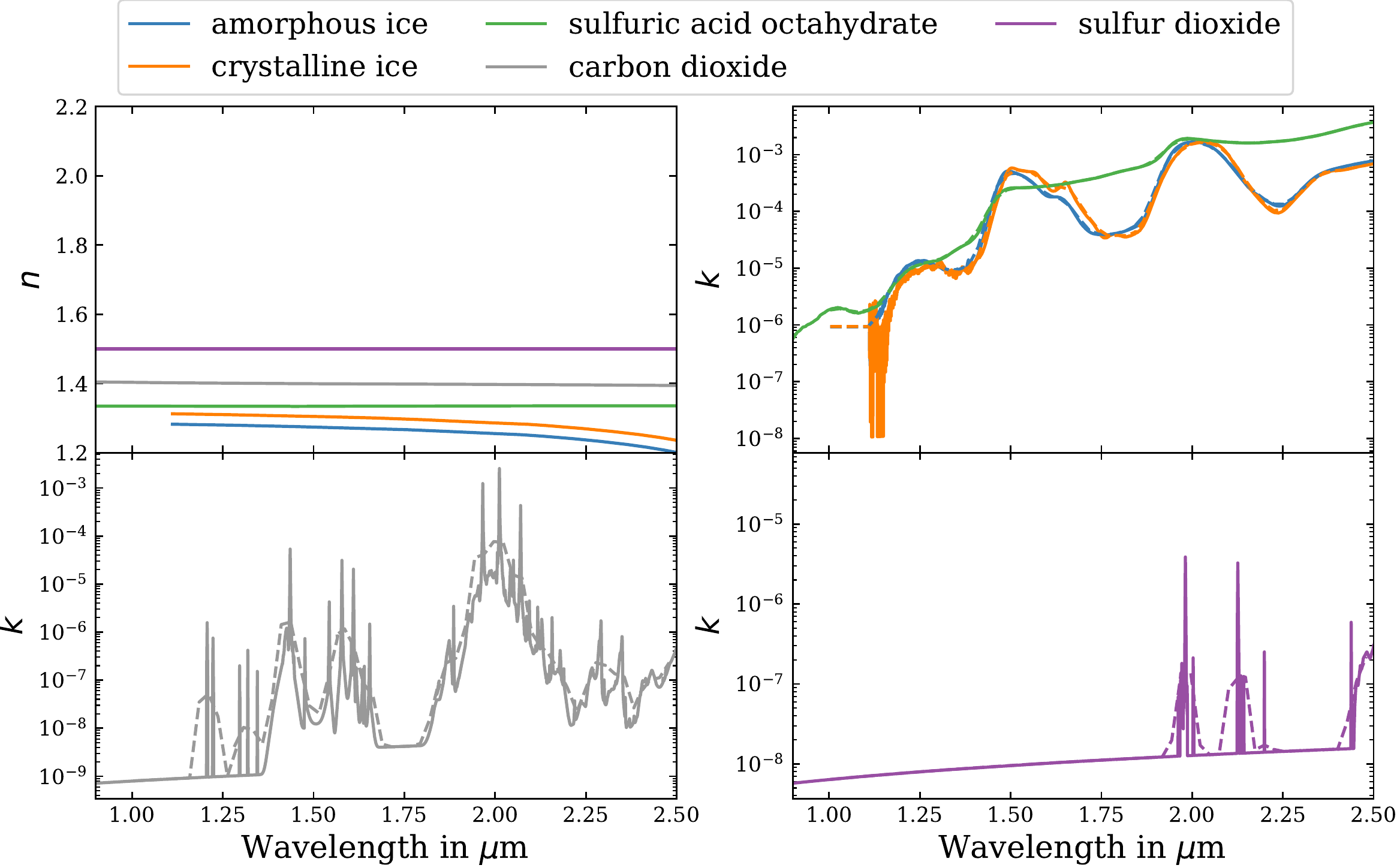}
\caption{Optical constants $n$ and $k$ of species included in this analysis. Dashed lines in the bottom-left, top-right and bottom-right panels show the $k$ spectra convolved to the NIMS resolution. CO$_2$ and SO$_2$ $k$ spectra have been individually shown in the bottom panels for clarity. Amorphous and crystalline ice optical constants, at 120 K, were published and included in \citet{mastrapa_optical_2009}; SAO optical constants at 77 K were measured by \citet{carlson_distribution_2005}; CO$_2$ optical constants, at 179 K, are from \citet{quirico_near-infrared_1997, quirico_composition_1999}; SO$_2$ optical constants, at 125 K, are from \citet{schmitt_identification_1994, schmitt_optical_1998}. CO$_2$ and SO$_2$ optical constants are available in the Solid Spectroscopy Hosting Architecture of Databases and Expertise or SSHADE\footnote{\url{sshade.eu}} \citep{sshade}. \label{fig:op_cons}}
\end{figure}

Finally, an important dataset needed for our model are the optical constants for the species considered, which are shown in Figure \ref{fig:op_cons} (sources are mentioned in the figure caption). We use amorphous and crystalline water-ice optical constants at 120~K, similar to \citet{carlson_distribution_2005}. For SAO, CO$_2$ and SO$_2$, we have selected available optical constants that were derived at temperatures close to the Europan temperature range of 80-130 K \citep{spencer_temperatures_1999}. We note that optical constants for the species considered here are temperature dependent, thus both the measured and theoretical reflectance spectra employed in this study will be sensitive to temperature as well \citep{dalton_spectroscopy_2010}. Future work that expands on either the species or wavelengths consider here would be best served by additional laboratory-derived optical constants obtained at Europan temperatures.

\section{Methods} \label{sec:methods}

We use a Bayesian inference framework to analyze $S1$, $S2$ and $S3$, which has been described in detail in \citet{mishra_bayesian_2021}. Briefly, the two major components of this framework are the \emph{Forward model} and the \emph{Bayesian posterior sampling}, described in the next two sections. The \emph{Bayesian posterior sampling} efficiently samples the parameter space we wish to explore, generating millions of instances of \emph{Forward model} spectra, and selects the samples for which the corresponding model instances fit the data well. From this collection of samples, one can get marginalized posterior distributions of individual parameters, pair-wise 2D distributions of parameters that highlight correlations, and a marginal probability of the model itself, also known as the \textit{Bayesian evidence}, that is used for model comparison \citep[e.g.][]{trotta_bayesian_2017}.

    
\subsection{Forward model} \label{sec:forward_model}

To a good approximation, the bidirectional reflectance of a planetary regolith \citep{hapke_2012} can be written as:

\begin{equation} \label{eq:hapke_RT}
   r_F(\mu,\mu_0,g, \lambda) = K \dfrac{\omega(\lambda)}{4}\dfrac{\mu_0}{(\mu + \mu_0)}[P(g,\lambda) + H(\omega,\mu/K)H(\omega,\mu_0/K) - 1]
\end{equation}

Here 
 
 \begin{itemize}
    \item $r_F$ is the radiance factor, which is the ratio of bidirectional reflectance of a surface to that of a perfectly diffuse surface (Lambertian) illuminated and observed at an incidence angle, $i=0$, relative to the surface normal. 
    \item $\mu$ is the cosine of the emergence angle $e$.
    \item $\mu_0$ is the cosine of the incidence angle $i$.
    \item $g$ is the phase angle.
    \item $K$ is the porosity coefficient.
    \item $\omega$ is the single scattering albedo.
    \item $P$ is the particle phase function.
    \item $H$ is the Ambartsumian-Chandrasekhar function that accounts for multiple scattered component of the reflection \citep{chandrasekhar_radiative_1960}.
 \end{itemize}
 
Here, we have ignored parameters related to backscattering and other opposition effects as they become important below small phase angles for the coherent backstatter effect and typically less than a few tens of degrees for the shadow-hiding opposition effect \citep{hapke_opposition_2012,Hapke_2021_opposition}, whereas the phase angles of all three NIMS spectra used here are much higher ($\sim30 \degree$, see Table \ref{tab:data}). In addition, we ignore the photometric contribution of macroscopic roughness, which Hapke characterizes with a mean topographic slope angle, $\overline{\theta}$. Our available range of photometric geometries for $S1$, $S2$, and $S3$ is too limited to constrain $\overline{\theta}$, which, even at a single phase angle, requires sufficient coverage of incidence and emission angles to characterize the limb-darkening behavior of the surface. Moreover, global average values of $\overline{\theta}$ derived from Voyager and Galileo imaging data vary little from $16^\circ$ to $22^\circ$ among different works \citep{verbiscer_photometric_2013}. At the photometric geometry of the NIMS spectra for S1, S2, and S3, the effects should be small enough to be ignored. However, we note that the detectability of photometric roughness is albedo-dependent such that the measurable values of $\overline{\theta}$ diminish as the surface albedo approaches unity because of multiply-reflected light among topographic surface facets. Since the particle spectral albedos vary with wavelength, it is reasonable to expect that there may be a wavelength-dependent variation in the effects of macroscopic roughness on our spectra. None of our reflectance values approach close enough to unity for this to be a significant concern.
 
The equations for the single scattering albedo, $\omega$, come from the equivalent-slab approximation \citep{hapke_single-particle_2012}. For the phase function $P$, we use a two-parameter \textit{Henyey-Greenstein} function \citep{henyey_diffuse_1941}, which has been shown to be representative of a wide variety of planetary regoliths, including the icy particles on Europa \citep{hapke_single-particle_2012}. A detailed description of all the parameters in eq. \ref{eq:hapke_RT} can be found in the Appendix of \citet{mishra_bayesian_2021}. 

We are also assuming that the regolith in the heavily bombarded and sputtered trailing hemisphere of Europa, where $S1$, $S2$ and $S3$ come from, is well-mixed or `churned' \citep{carlson_europas_2009, shirley_europas_2016}. Hence, particles of different species are mixed homogenously together in a intimate mixture. The averaging process in this \textit{intimate mixing model} is over the individual particle, and $\omega$ and $P$ in Eq. \ref{eq:hapke_RT} become volume averages of different materials in the mixture:

\begin{gather} \label{eq:im_eqn}
    \omega_{mix} = \dfrac{\sum_j N_j \sigma_j Q_{Ej} \omega_j}{\sum_j N_j \sigma_j Q_{Ej}} = \dfrac{\sum_j f_j \sigma_j Q_{Ej} \omega_j}{\sum_j f_j \sigma_j Q_{Ej}} \\
    P_{mix} = \dfrac{\sum_j N_j \sigma_j Q_{Ej} \omega_j P_j}{\sum_j N_j \sigma_j Q_{Ej} \omega_j} = \dfrac{\sum_j f_j \sigma_j Q_{Ej} \omega_j P_j}{\sum_j f_j \sigma_j Q_{Ej} \omega_j}
\end{gather}

where for a species of type $j$, $N_j$ is the number of particles per unit volume, $\sigma_j (= \pi D^2/4)$ is the cross-sectional area, $Q_{E,j}$ is the volume average extinction efficiency, $\omega_j$ is the the single scattering albedo, $P_j$ is the phase function, and $f_j$ is the number density fraction, equivalent to $N_j/\sum N_j$. For more details about the mixing equation parameters, please refer to section 2.2.2 of \citet{mishra_bayesian_2021}.

\subsection{Bayesian posterior sampling} \label{sec:Bayes_sampling}

The parameters we are fitting for, which define the parameter space are:

\begin{itemize}
    \item log$_{10}$($X_i$), where $X_i$ is the abundance (or number density fraction) of each species
    \item log$_{10}$($D_i$), where $D_i$ is the grain-size of each species
    \item filling factor $\phi$, which is the volume fraction occupied by particles, equivalent to 1 - (porosity of the regolith). $\phi$ is related to the porosity coefficient $K$ in eq. \ref{eq:hapke_RT} through the relation $K = -ln(1 - 1.209\phi^{2/3})/(1.209\phi^{2/3})$.
    \item a multiplicative factor $c$ for the model spectrum. In the 1-2.5 $\micron$ region, the absolute calibration of the NIMS instrument, or the accuracy of its measurements, is uncertain to around $\pm 10\%$ \citep{carlson_near-infrared_1992, carlson_distribution_2005}. 
    \item a wavelength-shift parameter $\Delta_{wav}$, to shift the absorption band centers of the model spectrum by $\Delta_{wav}$ , mimicking radiation induced shifts.
\end{itemize}

\begin{figure}
\centering
\includegraphics[width=\textwidth]{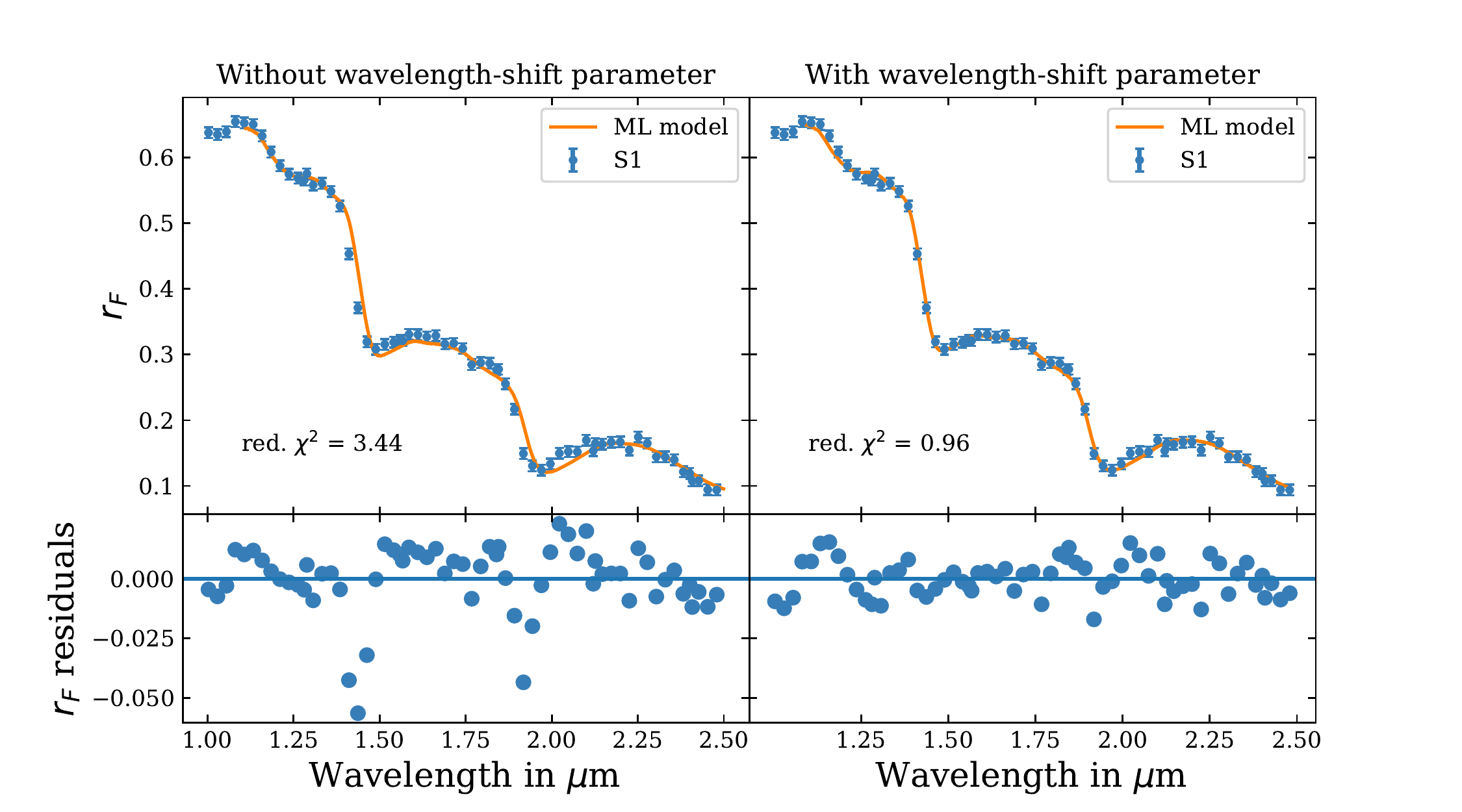}
\caption{Best-fit or maximum likelihood solutions of a SAO and amorphous ice model fit to $S1$ (blue points). The left panel shows the solution (orange) for the case where a wavelength-shift parameter is not included, while the right panel shows the solution (orange) for the case where that parameter is included (this turns out to be the preferred model for $S1$ as shown in section \ref{sec:results}). In the left panel, the model is slightly shifted to longer wavelengths as compared to the data, around the band-edges near 1.4 and 1.9 $\micron$. The reduced $\chi^2$ values of the fits are also noted, which illustrate that the fit in the right panel is better. \label{fig:radiation_shift}}
\end{figure}

We have included the $\Delta_{wav}$ parameter because Europa's trailing hemisphere spectra, like $S1$, $S2$ and $S3$, have water absorption-band centers that are shifted to slightly shorter wavelengths, by about tens of nanometers. This is a known effect of particle irradiation on fundamental absorption bands of hydrates, sulfates and silicates \citep[e.g.][]{dybwad_radiation_1971, nash_ios_1977}. For Europa specifically, it has been emphasized that comparisons of lab and numerically modelled spectra to Europa spectra must allow for the radiation-induced shifts of hydrate bands, so as to not bias our inference of the chemical composition from the spectra \citep{carlson_europas_2009}. One way to circumvent the problem would be to discard the data points near band-edges \citep[e.g][]{carlson_distribution_2005}. However, we find that our wavelength-shift parameter $\Delta_{wav}$ works sufficiently well and results in excellent fit to the data. An example is shown in Figure \ref{fig:radiation_shift}, where $S1$ is fit with a model with and without the $\Delta_{wav}$ parameter, with a much superior fit in the latter case. Moreover, measuring this wavelength shift for different spectra can also inform us about  trends in the level of irradiation experienced by the regoliths corresponding to the spectra. Hence, we include this parameter in all subsequent analysis.

To sample  this multi-dimensional parameter space, we use the \textit{nested sampling} algorithm \citep{skilling_nested_2006}. Specifically, we employ the Python implementation called \texttt{dynesty\footnote{\url{dynesty.readthedocs.io}}}, which implements dynamic nested sampling \citep{higson_dynamic_2019}, a more computationally accurate version of the standard nested sampling algorithm. Schematically, the iterative parameter exploration by nested sampling proceeds as follows:

\begin{enumerate}
    \item A random instance of a set of parameters or a parameter vector is drawn from the parameter space, defined by the \textit{prior distribution function}. For abundances, we use a Dirichlet prior \citep[e.g.][]{benneke_atmospheric_2012, lapotre_probabilistic_2017}, which ensures that the abundance parameters being sampled satisfy the constraint that they should sum up to 1. For the rest of the parameters, which do not have any such constraints, we use a uniform distribution, which is relatively uninformative (as compared to a Gaussian prior, for example), leading the data to drive the solution. The prior function also serves the purpose of defining the boundary of the parameter space to explore. Table \ref{tab:priors} lists all the free parameters, their prior function type and bounds\footnote{It should be noted that while the $\pm 10\%$ NIMS calibration uncertainty translates to a prior range of (0.9,1.1) for $c$, we ended up setting the lower limit to 0.7. The fits for $S1$ and $S2$ are very poor when the lower limit for $c$ is 0.9 and the derived posterior distribution of $c$ ‘hits the wall’ on the lower limit of 0.9, indicating that the model’s reflectance values are unable to achieve values as low as the data. One reason for the discrepancy could be that the in-flight calibration error is greater than the 10\% value found in the laboratory \citep{carlson_near-infrared_1992}. Secondly, irradiation has also been seen to reduce reflectance level and water band depths \citep{nash_ios_1977}. If the latter effect is at play here, then it will be folded into the $c$ parameter.}.
    \item Using these parameters, with the optical constants of the component species and the observation geometry parameters as the main inputs, the forward model (section \ref{sec:forward_model}) generates a model spectrum and then convolves it with the instrument response function (or bins it to the instrument’s resolution) to produce simulated data points. Each NIMS channel has a triangular response function \citep{carlson_near-infrared_1992, carlson_distribution_2005}, spanning the channel-width of $\sim 0.026 \ \micron$.
    \item These predicted data points are compared with observed data points to compute the posterior probability of this particular instance of parameters, defined by the \textit{Bayes' theorem} as follows:
    
    \begin{gather} \label{eq:bayes_theorem}
    p(\theta_j|d,M) = \dfrac{p(d|\theta_j,M) p(\theta_j|M)}{\int p(d|\theta_j,M) p(\theta_j|M) d\theta_j} \equiv \dfrac{\Lagr(d|\theta_j,M) \pi (\theta_j|M)}{\Z(d|M)}
    \end{gather}
    
    where $\Lagr (d| \theta, M_i)$ is the likelihood probability function, $\pi$ is the prior, $\Z$ is the Bayesian evidence and $p(\theta|d,M)$ is the posterior probability function. Since we assume the errors on our data to be Gaussian and independent, the likelihood function is also a Gaussian equal to 
    
    \begin{gather} \label{eq:likelihood}
    \Lagr (d| \theta, M_i) = {\displaystyle \prod^{N_{obs}}_{k=1}} \dfrac{1}{\sqrt{2\pi \sigma_k^2}} \textrm{exp} \Big(- \dfrac{\chi^2}{2}\Big)
    \end{gather}

    where $N_{obs}$ is the number of observed data points (i.e., number of wavelength channels/data points in the observed spectrum), $\chi^2$ is the familiar goodness-of-fit metric, and  $\sigma$ is the standard deviation. 
    
    \item  The posterior probability decides whether this particular parameter set is saved or selected and, in turn, informs the choice of the next parameter set as well. Details about this process/algorithm can be found in \citet{higson_dynamic_2019}. 
\end{enumerate}

\begin{deluxetable*}{cccc}[htbp!]
\tablenum{2}
\tablecaption{The priors used for all model parameters in our fitting analysis} \label{tab:priors}
\tablehead{
\colhead{Parameter} & \colhead{Description} & \colhead{Prior} & \colhead{Range}}
\startdata
$f_i$ & Number-density fraction or abundance & Dirichlet & 10$^{-3.0}$ to 1 \\
$D_i$ & Average grain diameter (microns) & log-uniform & 10 to 1000 \\
$\phi$ & filling factor & uniform & 0.01 to 0.52 \\
$c$ & NIMS calibration uncertainty parameter & uniform & 0.7 to 1.1 \\
$\Delta_{wav}$ & wavelength-shift parameter ($\mu$m) & uniform & -0.1 to 0.1 \\
\enddata
\tablecomments{$\phi$'s lower limit has been set to 0.01 because $K$, which is the model parameter dependent on $\phi$ (see eq. \ref{eq:hapke_RT}), plateaus at a value of 1 for $\phi \lesssim 0.01$. The upper limit of 0.52 comes from \citet{hapke_bidirectional_2008} where it is described as a critical value above which the medium is too tightly packed, such that coherent effects become important and diffraction can’t be ignored.}
\end{deluxetable*}

After this process converges, the final set of samples can be marginalized to get individual or pair-wise parameter distributions. Another useful quantity that the nested sampling process returns is the \textit{Bayesian evidence of the model} or $\Z$ (eq. \ref{eq:bayes_theorem}), which is discussed in the next section. It quantifies the probability of a model given the data, integrated over the entire parameter space (eq. \ref{eq:bayes_theorem}). Hence, two models can be readily compared using $\Z$, in a process called \textit{Bayesian model comparison}. We adapt this into a \textit{nested Bayesian model comparison} process, where we compare the evidence for a model with all species (the full model) with models where each species is removed one at a time. Assuming no \textit{a priori} preference of either the full model or the model without species \textit{X}, if the Bayesian evidence is higher for the latter, we conclude that \textit{X} is not needed to explain the data and hence there is no evidence for it. This process helps us select the simplest model, i.e., with the fewest number of species, compatible with the observations. The comparisons are also used to quantify the confidence in the presence of each species remaining in the simplest model, through the $\sigma$-significance metric (see section 2.2.3 of \citet{mishra_bayesian_2021} for a full description of statistical aspects of the Bayesian inference methodology). As per empirically calibrated evidence thresholds known as Jeffrey's scale \citep{jeffreys_1939}, a detection significance exceeding around 2.1$\sigma$, 2.7$\sigma$ and 3.6$\sigma$ levels are referred to as `weak', `moderate' and `strong' evidence respectively. We use these metrics to guide our interpretation of our model results. 

Over the course of analyzing $S1$, $S2$ and $S3$, we performed around 25 Bayesian inference analyses with models containing different combinations of the five species - amorphous ice, crystalline ice, SAO, CO$_2$ and SO$_2$. A full-model analysis (with all the five species) amounts to 12 free parameters and computes $\sim 4 \ \textrm{x} \ 10^5$ model spectra. Using this methodology, for each of the three spectra, we now proceed to describe the detection significance of all five species, the  simplest model that best explains the spectrum, and the constraints on the parameters of that model.




\section{Results} \label{sec:results}

The results for the nested model comparisons performed for $S1$, $S2$ and $S3$ are shown in Table \ref{tab:results_bmc}. For each spectrum, we first perform a Bayesian inference analysis with a full model, i.e., with all five species - amorphous ice, crystalline ice, SAO, CO$_2$ and SO$_2$. We then remove each species one at a time, and rerun the analysis with these alternative models. Comparing each of their results against the reference model with all five species, we can quantify the evidence for the presence of each species. The retrieved parameter constraints for the models with highest evidence (i.e., the simplest models that explain the data) are shown in Table \ref{tab:results_params}.

\begin{deluxetable*}{lcccccc}[htbp!]
\tablenum{3}
\tablecaption{Results of nested model comparisons for $S1$, $S2$ and $S3$ \label{tab:results_bmc}}
\tablewidth{0pt}
\tablehead{
\colhead{} & \multicolumn{2}{c}{$S1$} & \multicolumn{2}{c}{$S2$} & \multicolumn{2}{c}{$S3$} \\
\colhead{Model} & \colhead{ln($\Z_i$)} & \colhead{$\chi^2_{r,min}$} & \colhead{ln($\Z_i$)} & \colhead{$\chi^2_{r,min}$} & \colhead{ln($\Z_i$)} & \colhead{$\chi^2_{r,min}$}}
\startdata
All species (ref.) & 200.11 & 1.03 & 210.831 & 1.02 & 236.85 & 0.73  \\
No SAO & -2453.90 ($>30\sigma$) & 99.12 & -2523.39 ($>30\sigma$) & 102.17 & -423.48 ($>30\sigma$) & 25.22 \\
No amorphous ice & 198.45 (2.4$\sigma$) & 1.04 & 211.01 (N.A.) & 1.1 & 232.67 ($3.36\sigma$) &  0.96 \\
No crystalline ice & 200.87 (N.A.) & 0.99 & 194.96 ($5.9\sigma$) & 1.69  & 227.28 ($4.76\sigma$) & 1.17 \\
No CO$_2$ & 200.48 (N.A.) & 0.99 & 212.17 (N.A.) & 0.98 & 238.21 (N.A.) & 0.70 \\
No SO$_2$ & 200.55 (N.A) & 1.02 & 212.36 (N.A) & 0.98 & 237.2 (N.A.) & 0.71 \\
\enddata
\tablecomments{For each of the three spectra, $S1$, $S2$ and $S3$, the Bayesian evidence ln($\Z_i$) and the minimum reduced chi-squared $\chi^2_{r,min}$, i.e, chi-squared of the maximum-likelihood solution are listed for all models. The `All species' model is the reference model to compare to and contains amorphous water-ice, crystalline water-ice, SAO, \COtwo{} and \SOtwo{}. The other four alternative models are each without one of the five species as specified. In brackets next to the ln($\Z_i$) values, an $n$-$\sigma$ detection indicates the degree of preference for the reference model over the alternative model, which can also be interpreted as the significance of detection of the species excluded from the alternative model. `N.A.' indicates that no evidence for the particular species has been found.}
\end{deluxetable*}

\begin{deluxetable*}{lccc}[htbp!]
\tablenum{4}
\tablecaption{Derived parameter values, along with the 1-$\sigma$ upper and lower bounds (68\% confidence intervals), for models with highest evidence for $S1$, $S2$ and $S3$. The parameters are described in Table \ref{tab:priors}. The subscripts `sao', `am' and `cr' stand for sulfuric-acid-octahydrate (SAO), amorphous ice and crystalline ice, respectively. The values in the brackets are the exponentiated versions of the logarithmic results. \label{tab:results_params}}
\tablewidth{0pt}
\tablehead{
\colhead{Parameter} & \colhead{$S1$} & \colhead{$S2$} & \colhead{$S3$}}
\startdata
log$_{10}f_{sao}$ ($f_{sao}$) & -0.0599$^{+0.0348}_{-0.0942}$ (0.8711$^{+0.0727}_{-0.1698}$) &  -0.0010$^{+0.0003}_{-0.0005}$  (0.9976$^{+0.0007}_{-0.0012}$) & -0.1478$^{+0.0771}_{-0.1136}$ (0.7115$^{+0.1382}_{-0.1638}$) \\
log$_{10}D_{sao}$ ($D_{sao}$) & 1.6254$^{+0.0576}_{-0.0502}$  (42.2071$^{+5.9857}_{-4.6100}$) & 1.6530$^{+0.0301}_{-0.0312} ($44.9820$^{+3.2296}_{-3.1217}$) &  1.9905$^{+0.0497}_{-0.0506}$ (97.8381$^{+11.8737}_{-10.7525}$) \\
log$_{10}f_{am}$  ($f_{am}$) & -0.8897$^{+0.3649}_{-0.3605}$ (0.1289$^{+0.1698}_{-0.0727}$) & - & -0.5456$^{+0.1973}_{-0.2870}$ (0.2847$^{+0.1637}_{-0.1377}$)\\
log$_{10}D_{am}$ ($D_{am}$) & 1.6919$^{+0.1175}_{-0.1291}$ (49.1912$^{+15.2787}_{-12.6505}$) & - & 2.0233$^{+0.1065}_{-0.1232}$ (105.5145$^{+29.3287}_{-26.0577}$)\\
log$_{10}f_{cr}$ ($f_{cr}$) & - & -2.6273$^{+0.1822}_{-0.1643}$ (0.0024$^{+0.0012}_{-0.0007}$) & -2.4494$^{+0.1003}_{-0.0803}$ (0.0036$^{+0.0009}_{-0.0006}$)\\
log$_{10}D_{cr}$ ($D_{cr}$) & - & 2.5826$^{+0.0808}_{-0.0887}$ (382.4395$^{+78.1922}_{-70.6589}$) & 2.9634$^{+0.0267}_{-0.0484}$ (919.1472$^{+58.2593}_{-96.8802}$)\\
$\phi$ & 0.1511$^{+0.0771}_{-0.0669}$ & 0.1883$^{+0.0631}_{-0.0613}$ & 0.0319$^{+0.0357}_{-0.0161}$ \\
$c$ & 0.7898$^{+0.0034}_{-0.0034}$ &  0.8029$^{+0.0049}_{-0.0055}$ & 0.9887$^{+0.0080}_{-0.0095}$ \\
$\Delta_{wav}$ & -0.0198$^{+0.0015}_{-0.0016}$ & -0.0098$^{+0.0010}_{-0.0010}$ & -0.0043$^{+0.0015}_{-0.0015}$ \\
\enddata
\tablecomments{The units for $D_i$ are microns. A `-' indicates that the species was not part of the preferred model. CO$_2$ and SO$_2$ have not been included in this table as neither of them were detected in any of the three spectra.}
\end{deluxetable*}

\begin{figure}[htbp!]
\centering
\includegraphics[width=\textwidth]{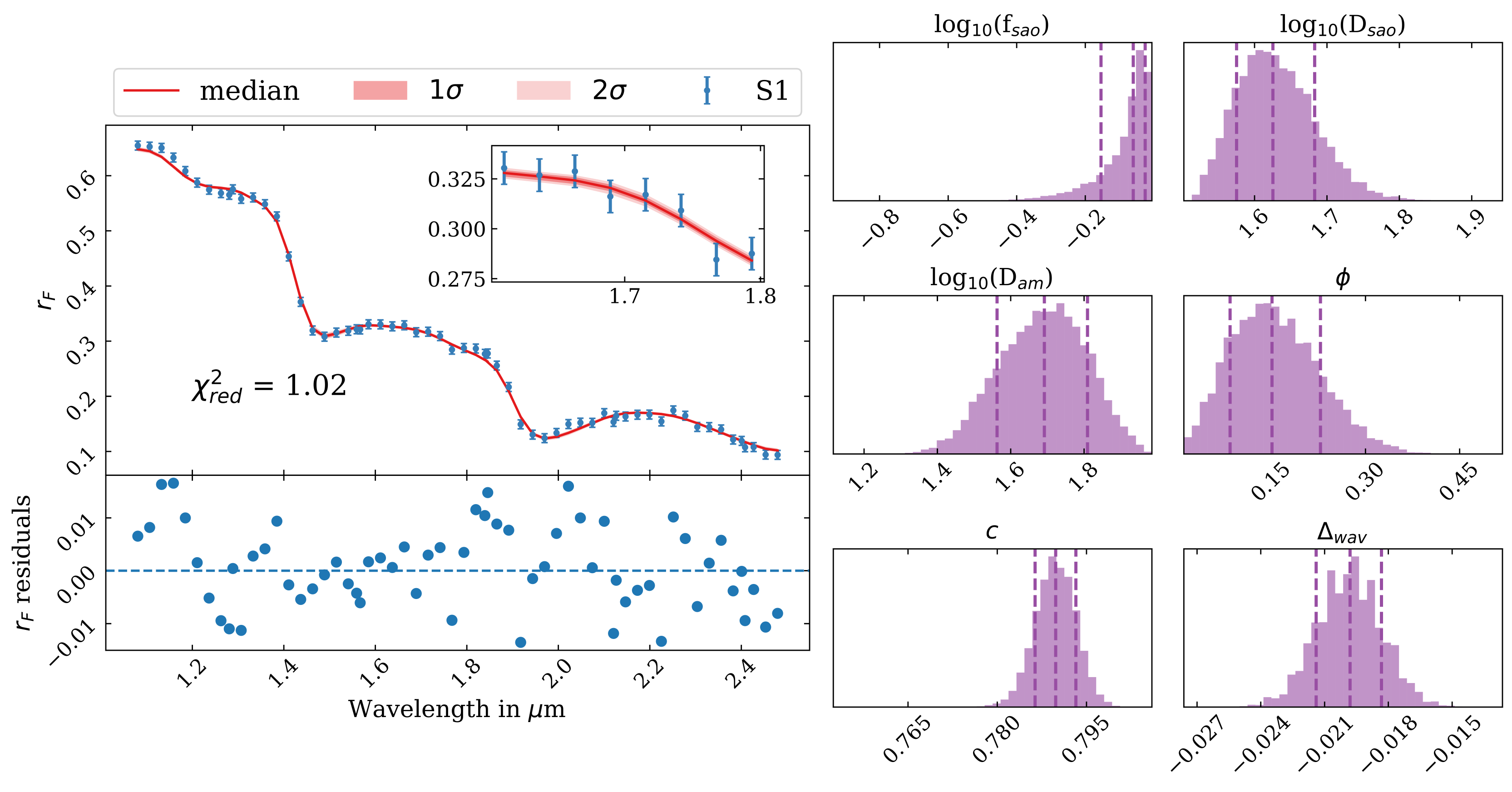}
\caption{Results from analysis of $S1$ with the SAO and amorphous ice mixture model. The top subplot in the left panel shows the median retrieved spectrum (red) for the fit to data (blue). The dark and light red regions indicate the $1\sigma$ and $2\sigma$ confidence regions around the median spectrum, better visible in the inset figure that zooms in around 1.7$\micron$. The reduced $\chi^2$ value of the fit of the median spectrum is 1.02 and the residuals to this fit are shown in the bottom subplot. The right panel shows the posterior distributions of the free parameters. The parameters are listed in Table \ref{tab:priors}. The dashed purple lines are the median values along with the 1-$\sigma$ upper and lower bounds (68\% confidence intervals). All the values are listed in Table \ref{tab:results_params}. Note that the distribution for $f_{am}$ has not been shown here as it is dependent on $f_{sao}$ and not one of the free parameters. The derived values for $f_{am}$ are shown in Table \ref{tab:results_params}. \label{fig:S1_results}}
\end{figure}

For $S1$, we find that there is a very strong evidence for SAO (at $> 30\sigma$ confidence level). This is expected as the octahydrate, which is taken to generically represent hydrates in our model, is the dominant component in the trailing hemisphere of Europa, as established in previous studies \citep[e.g.][]{carlson_distribution_2005, brown_salts_2013, ligier_vlt/sinfoni_2016}. Weak evidence for amorphous ice at a $2.4\sigma$ confidence-level is also found. We don't find any evidence for crystalline ice, CO$_2$ or SO$_2$ , which means that the simplest model that explains $S1$ consists of only SAO and amorphous ice. Indeed, a nested model comparison with the SAO and amorphous ice mixture as the reference model (not shown in Table \ref{tab:results_bmc}) provides very strong evidence for both: $>30\sigma$ for SAO and $8.6\sigma$ for amorphous ice. Detailed results from the analysis with the SAO and amorphous ice mixture model are presented in Figure \ref{fig:S1_results}. The median retrieved spectrum provides an excellent fit to $S1$, with a reduced $\chi^2$ value of 1.02. Also shown in the figure are the posterior probability distributions of the model parameters. SAO dominates the composition, with its abundance constrained as $87.11_{-16.98}^{+7.27}\%$ (median value with lower and upper 1$\sigma$ limits) while amorphous ice is present in a small amount of $12.89_{-7.27}^{+16.98}\%$. Both SAO and amorphous ice have small average grain-sizes, of $42.21_{-4.61}^{+5.98}$ microns and $49.19_{-12.65}^{+15.28}$ respectively. The regolith is fairly porous, with a tightly constrained filling factor $\phi$ of 0.15$_{-0.06}^{+0.07}$ (porosity is defined as $1-\phi$). The wavelength shift parameter $\Delta_{wav}$ has also been well constrained to -0.0198$_{-0.0016}^{+0.0015}$, slightly less than the NIMS channel width of 0.026 $\micron$. This means that the water band centers at around $1.5$ and $2.0 \micron$ in the model spectrum needed to be shifted towards shorter wavelengths by around 0.02 $\micron$ or 20 nm to account for the radiation-induced shift seen in the data.

\begin{figure}[htbp!]
\centering
\includegraphics[width=\textwidth]{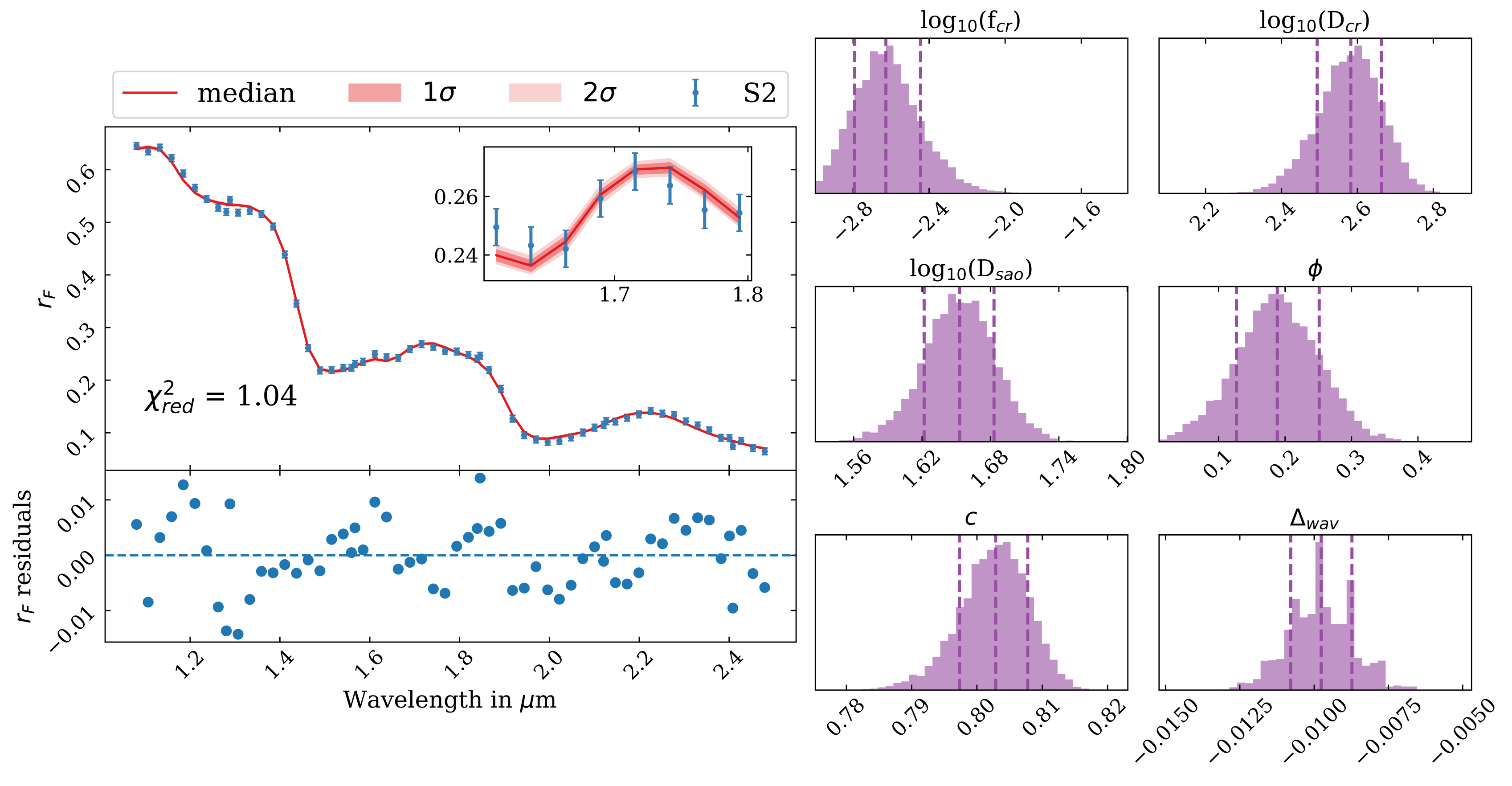}
\caption{Results from analysis of $S2$ with the SAO and crystalline ice mixture model, presented in the same form as Figure \ref{fig:S1_results}. Note that the distribution for $f_{sao}$ has not been shown here as it is dependent on $f_{cr}$ and not one of the free parameters. The derived values for $f_{sao}$ are shown in Table \ref{tab:results_params}. \label{fig:S2_results}}
\end{figure}

Next, for $S2$, we once again find very strong evidence ($>30\sigma$) for SAO. While no evidence for amorphous ice is found, $S2$ shows a strong (5.9$\sigma$) evidence for crystalline ice. No evidence for either CO$_2$ or SO$_2$ is found, which means that the simplest model that fits the data comprises of a mixture of SAO and crystalline ice. Nested model comparisons with SAO and crystalline ice mixture model as the reference provides very high evidence for both SAO ($>30\sigma$) and crystalline ice ($29.7\sigma$). The results of the analysis with this preferred model are shown in Figure \ref{fig:S2_results}. The median retrieved spectrum provides an excellent fit to the data, with a reduced $\chi^2$ value of 1.04. All the model parameters have well constrained Gaussian distributions. Crystalline ice is present in a very minor amount ($0.24_{-0.0007}^{+0.0012}\%$), but with very large grains of average size $382.44_{-70.66}^{+78.19}$ microns. SAO heavily dominates the composition with an abundance of $99.76_{-0.0012}^{+0.0007}\%$, but smaller grains of average size of $44.98_{-3.12}^{+3.23}$ microns. The filling factor $\phi$ is low (0.18$_{-0.06}^{+0.06}$), indicating a porous regolith. Finally, a slight band-center shift is once again detected and the wavelength-shift parameter $\Delta_{wav}$ is constrained to -0.0098$_{-0.001}^{+0.001}$, which is about half the channel-width of NIMS.

\begin{figure}[htbp!]
\centering
\includegraphics[width=\textwidth]{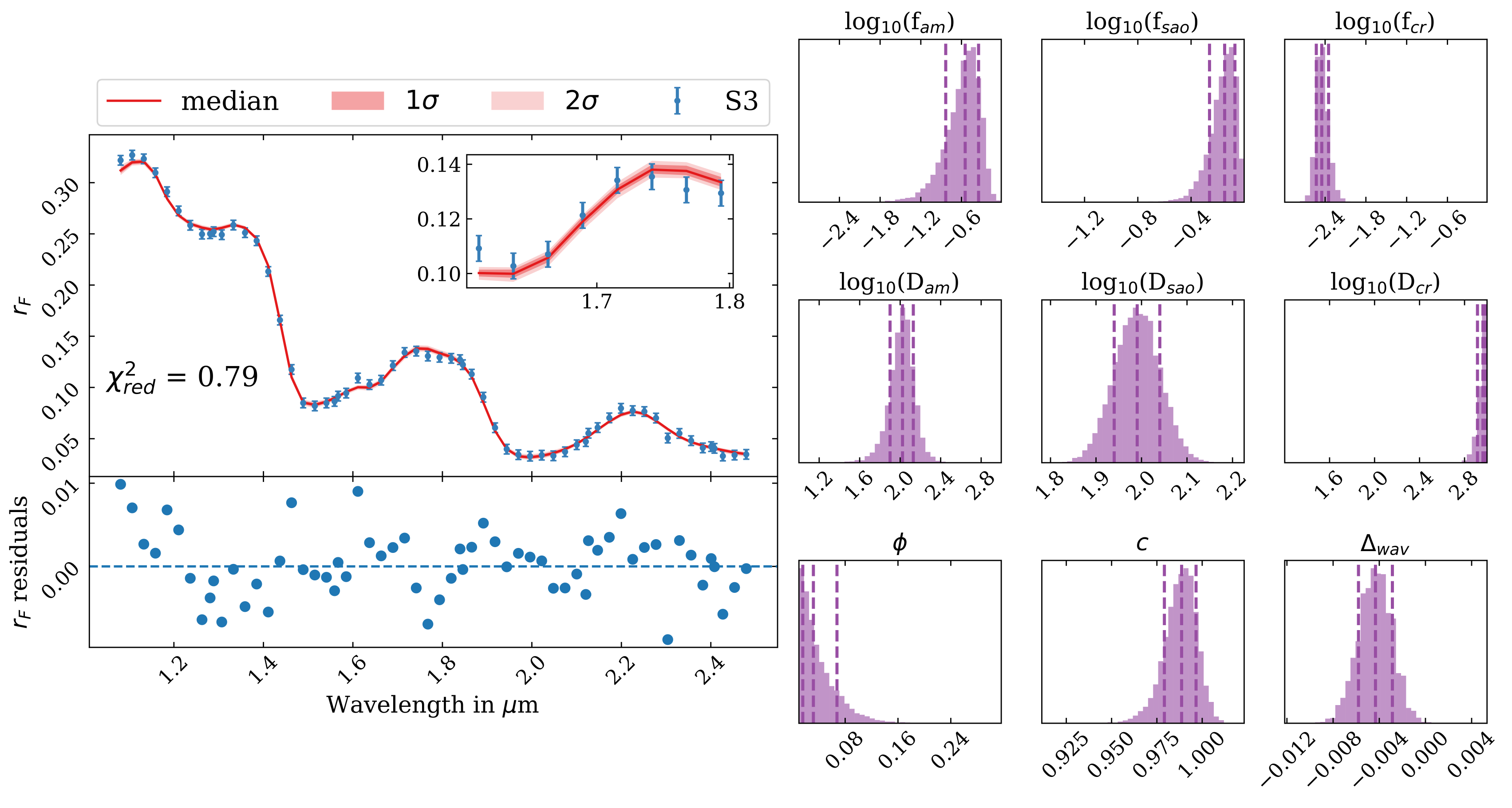}
\caption{Results from analysis of $S3$ with the SAO, amorphous ice and crystalline ice mixture model, presented in the same form as Figure \ref{fig:S1_results}.\label{fig:S3_results}}
\end{figure}

Finally, for $S3$, we find very strong evidence for SAO ($>30 \sigma$) and crystalline ice ($4.76\sigma$), whereas a moderate to strong evidence for amorphous ice ($3.36\sigma$). Once again, no evidence is found for either CO$_2$ or SO$_2$. The preferred model is thus a SAO, amorphous ice and crystalline ice mixture, whose results are shown in Figure \ref{fig:S3_results}. Nested comparisons for this model provides strong detection-levels for all three components: $>30\sigma$ for SAO, 3.8$\sigma$ for amorphous ice, and 7.2 $\sigma$ for crystalline ice. The median retrieved spectrum, as shown in Figure \ref{fig:S3_results}, provides a good fit to the data, with a reduced $\chi^2$ value of 0.79. The retrieved parameter distributions tell us that SAO is once again the dominant component ($71.15_{-16.38}^{+13.82}\%$), albeit with an abundance slightly lower as compared to the values for $S1$ and $S2$. Amorphous ice is also a major species, with an abundance of $28.47_{-13.77}^{+16.37}\%$, while crystalline ice is present in trace amounts of $0.36_{-0.0006}^{+0.0009}\%$. However, crystalline ice's contribution to the overall absorption comes from the very large average grain-size of $919.66_{-96.88}^{+58.88}$ microns. SAO and amorphous ice meanwhile have moderate average grain-size values of $97.83_{-10.75}^{+11.87}$ and $105.91_{-26.05}^{+29.32}$ microns respectively. The regolith is constrained to be porous once again, with a significantly low value for $\phi$ of (0.03$_{-0.01}^{+0.03}$). Finally, the spectral band-centers are negligibly shifted, as the $\Delta_{wav}$ parameter is retrieved to be -0.0043$_{-0.0015}^{+0.0015}$ $\micron$.

\vspace{5pt}

\begin{figure}[htbp!]
\centering
\includegraphics[width=\textwidth]{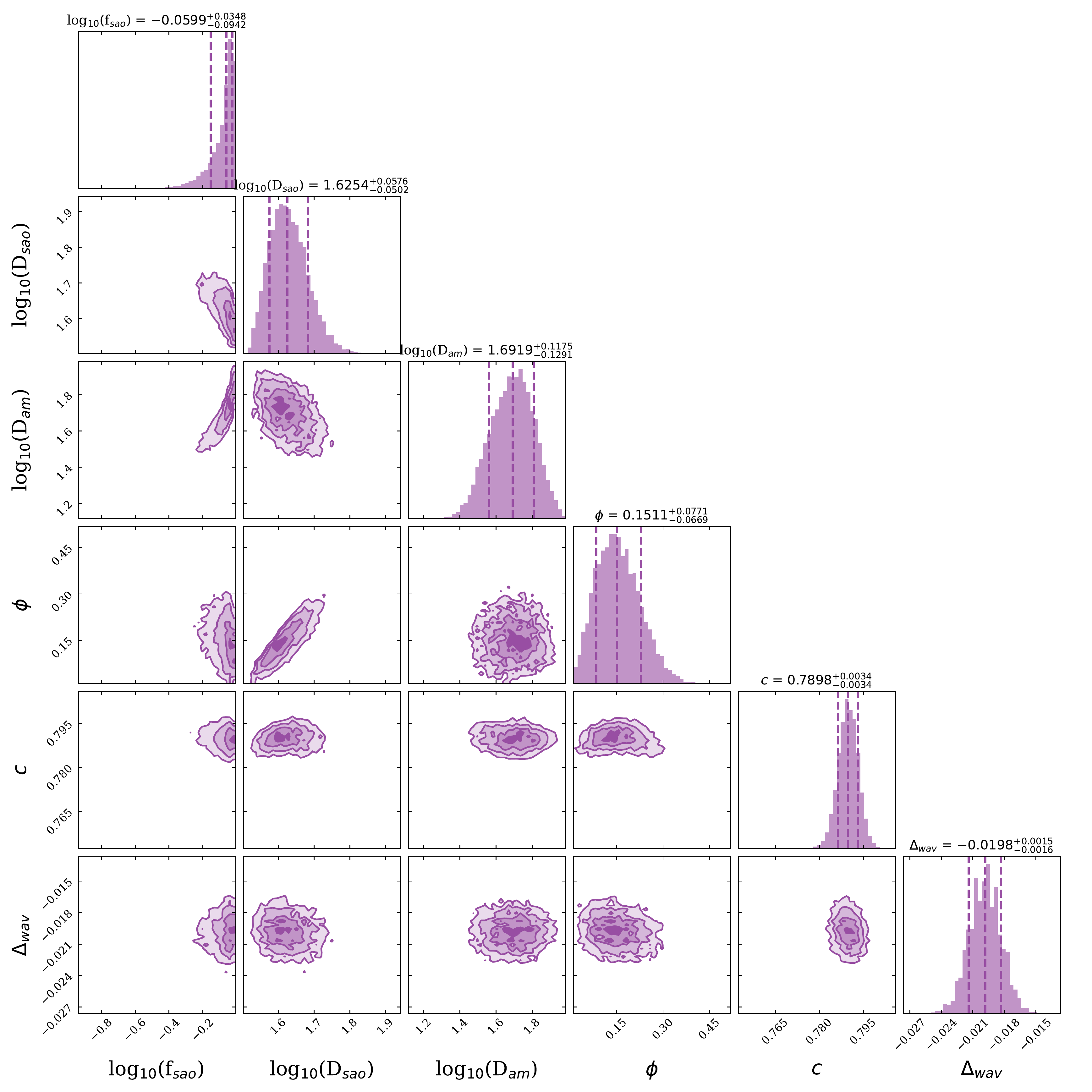}
\caption{Full `corner' plot the from analysis of $S1$ with the SAO and amorphous ice mixture model. The 1D histograms along the diagonal show marginalized posterior probability distributions of all the individual parameters. The dashed purple lines are 1-$\sigma$ upper and lower bounds (68\% confidence intervals). Each diagonal plot also shows the corresponding median values and the associated 1$\sigma$ lower and upper limits for the distribution. The remaining plots are pair-wise 2D distributions, that illustrate the correlations between parameters. The contours in these 2D distributions correspond to 0.5,1,1.5 and 2.0$\sigma$ intervals. Parameter symbols are described in Table \ref{tab:priors}. \label{fig:S1_corner}}
\end{figure}

\begin{figure}[htbp!]
\centering
\includegraphics[width=\textwidth]{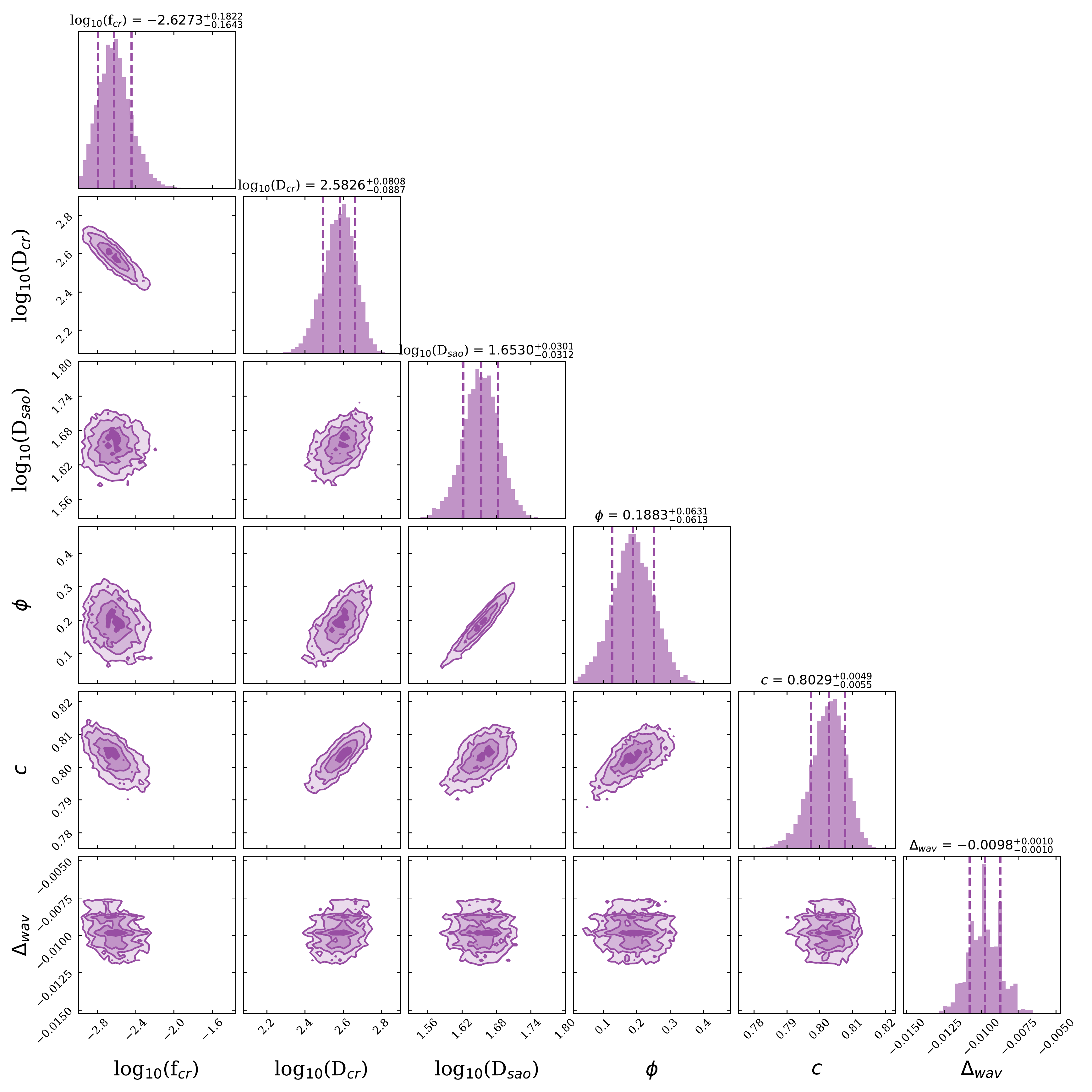}
\caption{Full `corner' plot the from analysis of $S2$ with the SAO and crystalline ice mixture model, presented in the same form as Figure \ref{fig:S1_corner}. \label{fig:S2_corner}}
\end{figure}

\begin{figure}[htbp!]
\centering
\includegraphics[width=\textwidth]{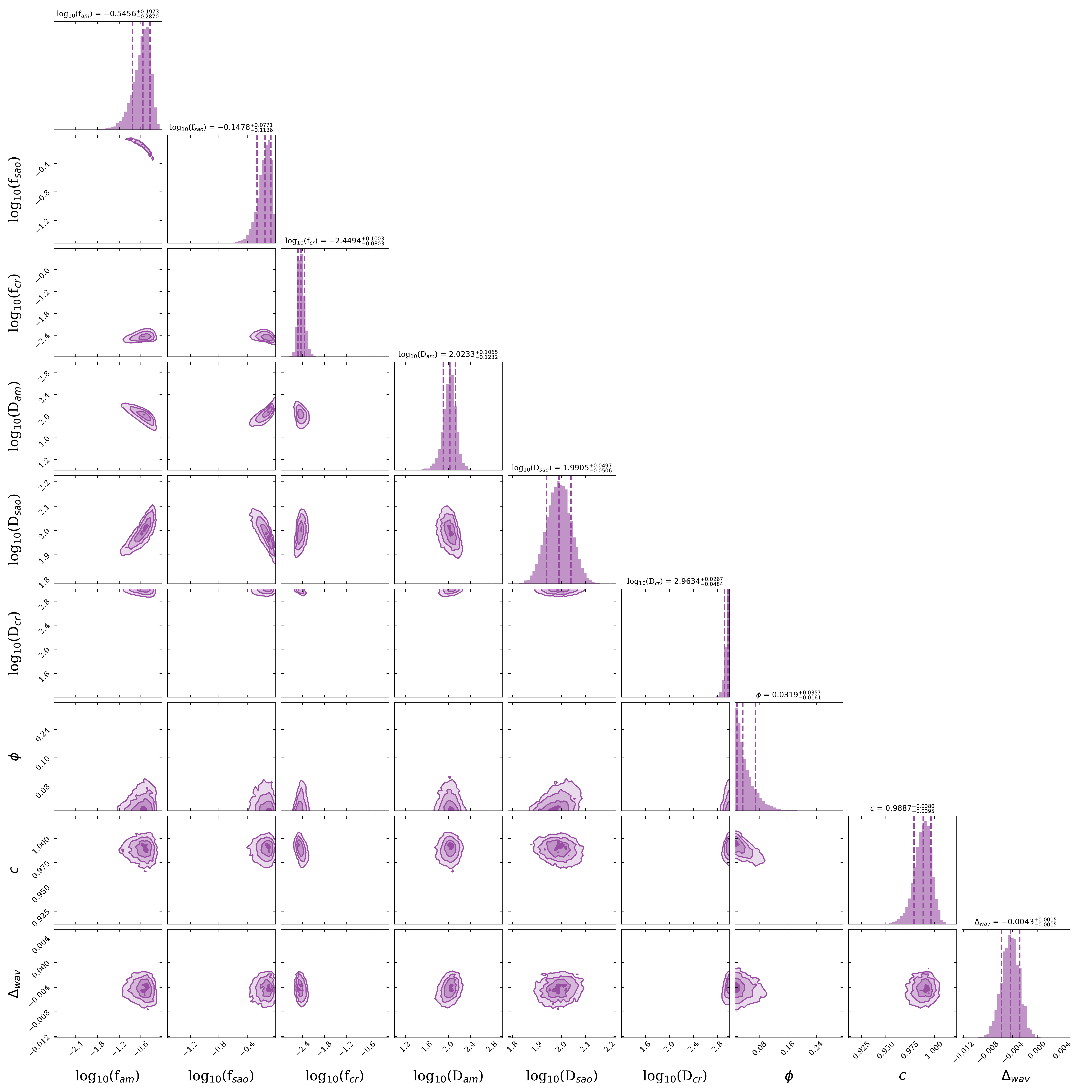}
\caption{Full `corner' plot the from analysis of $S3$ with the SAO, amorphous ice and crystalline ice mixture model, presented in the same form as Figure \ref{fig:S1_corner}. \label{fig:S3_corner}}
\end{figure}

Along with the individual parameter distributions that are shown in Figures \ref{fig:S1_results}-\ref{fig:S3_results}, our analysis also yields pair-wise distributions of the various parameters (Figures \ref{fig:S1_corner}-\ref{fig:S3_corner}), which highlight the correlations between them. Some correlations are expected, such as the correlation between grain-size and abundances. In all three cases, we see that the grain-size of any species (SAO, amorphous ice or crystalline ice) is anti-correlated with its abundance. This makes sense as the `weight' of a species in the mixing equation in Hapke's model is proportional to the product of its abundance and its grain-size squared. Hence, decreasing the abundance of a species should have the same overall effect as increasing its grain-size. 

On the other hand, Figures \ref{fig:S1_corner}-\ref{fig:S3_corner} also illuminate complex parameter correlations, e.g., the correlation between the filling factor $\phi$ and the calibration parameter $c$ is puzzling - we see that $\phi$ and $c$ show no correlation for $S1$, positive correlation for $S2$ and negative correlation for $S3$. The filling factor $\phi$ enters the Hapke radiative transfer equation (eq. \ref{eq:hapke_RT}) as the porosity coefficient $K$, via the relation $K = -ln(1 - 1.209\phi^{2/3})/(1.209\phi^{2/3})$. $K$ affects eq. \ref{eq:hapke_RT} as a multiplicative factor and via the multiple scattering function $H$. When $\phi$ and $c$ are allowed to vary in isolation (the abundance and grain-size parameters are kept fixed), they are highly anti-correlated, since both essentially act like multiplicative factors. However, in our exercise where abundance and grain-size parameters are free as well, $\phi$ is not the only parameter affecting the multiple scattering function $H$, and hence the correlations between $\phi$ and $c$ become complex. 

\section{Establishing detectability limits for CO$_2$ and SO$_2$} \label{sec:detectability}

Since neither CO$_2$ nor SO$_2$ were detected in any of the three spectra, it was of interest to perform a theoretical study that establishes the minimum amount of CO$_2$ and SO$_2$ that can be detected with NIMS data, in the 1-2.5 $\micron$ range. Figure \ref{fig:op_cons} shows that in the 1-2.5 $\micron$ range, both CO$_2$ and SO$_2$ possess sharp features in their $k$ (imaginary part of refractive index) spectra. CO$_2$ especially has numerous features, with a few at around 2.0 microns matching the strength of water-ice features. Since $k$ is a proxy for the absorption coefficient ($\alpha=4\pi k/\lambda$), the sharp/broad nature and position of peaks in the $k$ spectrum also determine the absorption features in the reflectance spectrum. 

\begin{figure}[htbp!]
\centering
\includegraphics[width=\textwidth]{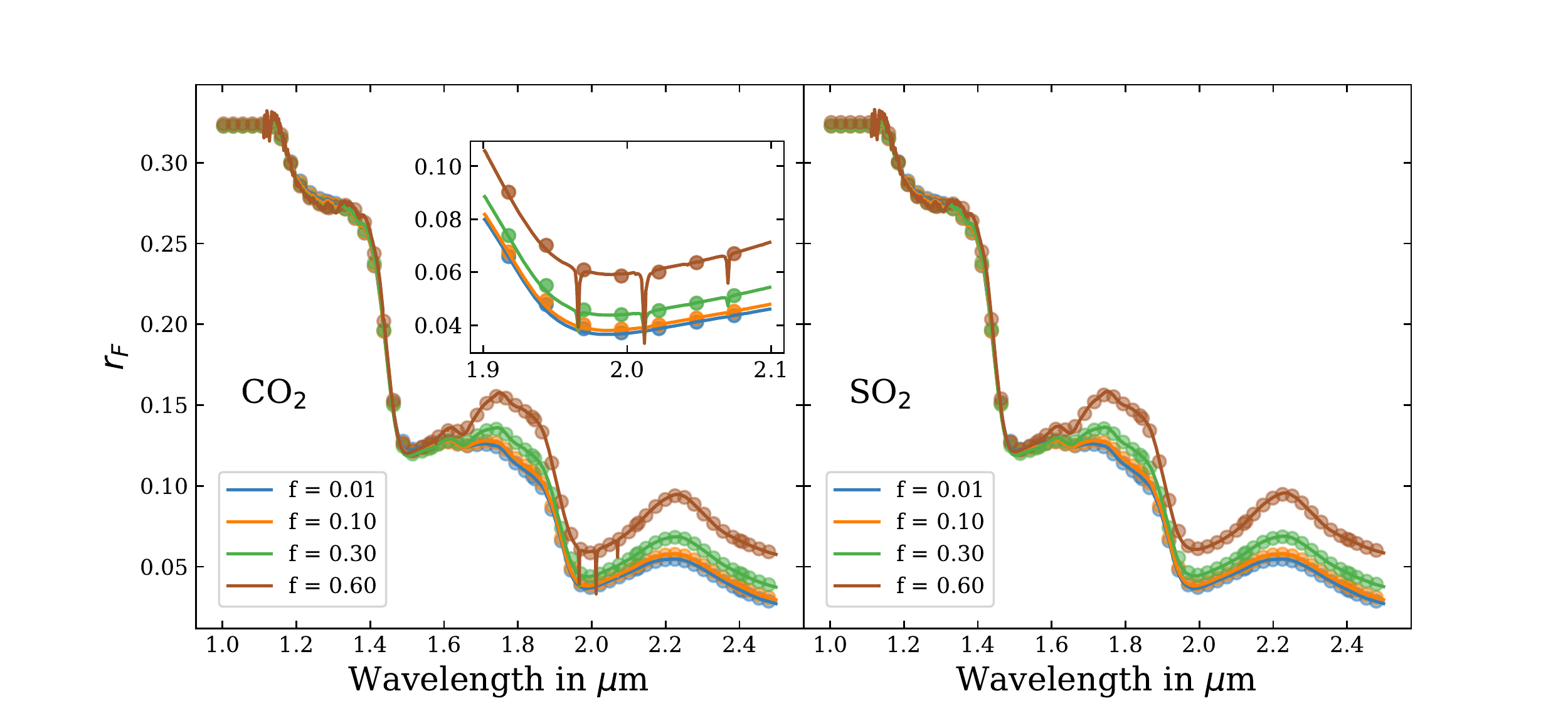}
\caption{Theoretical reflectance spectra of SAO, amorphous ice, crystalline ice and one of CO$_2$ (\textit{left}) or SO$_2$ (\textit{right}) are shown in the two panels. Starting with a SAO, amorphous ice and crystalline ice mixture model, with parameters derived from the $S3$ analysis (Figure \ref{fig:S3_results}), some amount of CO$_2$ (or $SO_2$) was added. SAO's abundance was reduced to account for the new species added. Solid lines show the high resolution model at the native resolution of the optical constants, of $\sim 0.001 \micron$. The circles show the models binned to the NIMS resolution of $\sim 0.026 \micron$. The inset subplot in the left panel zooms in around 2.0 $\micron$ where absorption features of CO$_2$ are prominent in the high-resolution model (brown) with an unrealistically high CO$_2$ abundance of 60\%, although no features are seen in the corresponding NIMS resolution model. No features of SO$_2$ appear in any of the cases considered here.  \label{fig:detection_limits}}
\end{figure}

We investigate the effect of CO$_2$ and SO$_2$ on the net reflectance in a mixture with the dominant species SAO, amorphous ice and crystalline ice also present. An example is shown in Figure \ref{fig:detection_limits}, where theoretical spectra of a mixture of SAO, amorphous ice, crystalline ice and CO$_2$/SO$_2$ were generated. We took the solution of the SAO, amorphous ice and crystalline ice mixture from the $S3$ analysis (Figure \ref{fig:S3_results}) and injected a fourth species - either CO$_2$ or SO$_2$ - in increasing amounts (SAO's abundance was being reduced to ensure the abundance fractions add up to 1). For CO$_2$, only at an unrealistically high abundance of around 60\% does its absorption features near 2.0 $\micron$ start to appear in the high-resolution spectrum (resolution $\sim 0.001 \micron$). These features do not appear in the coarser NIMS resolution spectrum (resolution $\sim 0.026 \micron$). For SO$_2$, no sharp features appear at native resolution even at abundance values approaching 60\%. Figure \ref{fig:detection_limits} also illustrates that both CO$_2$ and SO$_2$ affect the overall continuum level of the spectra as well. However, this effect is degenerate with parameters like the filling factor $\phi$, which also changes the overall amplitude of the spectrum. A Bayesian model comparison analysis shows that significant evidence ($>3\sigma$) for CO$_2$ and SO$_2$ only starts emerging at abundances of 60\% or higher. Hence, given the low expected abundance of around a few percent for CO$_2$ and SO$_2$ on Europa's surface \citep{hand_energy_2007, carlson_europas_2009}, and the relatively coarse spectral resolution of NIMS, detection of CO$_2$, SO$_2$ and possibly other oxides like H$_2$O$_2$ in the 1-2.5 $\micron$ wavelength range Galileo/NIMS spectra is unlikely. 

\section{Summary and Discussion} \label{sec:discussion}

We applied a novel Bayesian inference framework to three \emph{Galileo} NIMS observations of Europa (Figure \ref{fig:data}), referred to as $S1$, $S2$ and $S3$. Our analysis framework successfully evaluated evidence for and provided constraints on abundances and average grain-sizes of important trailing hemisphere species - amorphous water ice, crystalline ice, sulfuric-acid-octahydrate (SAO), CO$_2$, and SO$_2$. These parameter constraints account for the degeneracies with physical parameters of the regolith, like porosity and radiation-induced spectral band-center shift. We also employ Bayesian model comparison to find the simplest model that explains each spectrum, for which we also evaluate the statistical significance of detection of each species. The model comparisons (Table \ref{tab:results_bmc}) show that $S1$ is best explained by a mixture of SAO and amorphous ice, $S2$ by a mixture of SAO and crystalline ice, and $S3$ with a mixture of SAO, amorphous ice, and crystalline ice. The retrieved species abundances and grain-sizes for these preferred models are shown in Table \ref{tab:results_params}. No evidence for either CO$_2$ or SO$_2$ is found for any of the three spectra, but given the limitations of the Galileo NIMS data we cannot set useful limits on the abundances of these two species. Oxides, in trace amounts, can also be trapped in the matrix of the major species. Hence, the use of optical constants for CO$_2$ or SO$_2$ trapped in water-ice, rather than those of pure CO$_2$ or SO$_2$ ice, may be more appropriate. However, because of the minimal impact of CO$_2$ and SO$_2$ in the current dataset (see section \ref{sec:detectability} and Figure \ref{fig:detection_limits}), the choice of optical constants for the two oxides should have minimal impact on our main conclusions.

SAO dominates the composition ($>30\sigma$ detection) for all three spectra, with $S3$ having the lowest amount of SAO among the three. This is consistent with $S3$'s location at a high latitude of 67.9$\degree$, where the abundance of SAO is expected to be lower than nearer the equator \citep{carlson_distribution_2005, ligier_vlt/sinfoni_2016, brown_salts_2013}. \citet{carlson_distribution_2005} analyzed the same spectra and derived SAO's abundance to be 89\% for $S1$, 54\% for $S2$ and 30\% for $S3$. While the abundance we retrieve for $S1$, $87.11_{-16.98}^{+7.27}\%$, is consistent with their result to within 1-$\sigma$, our numbers for $S2$ ($99.76_{-0.0012}^{+0.0007}\%$) and $S3$ ($71.15_{-16.38}^{+13.82}\%$) do not agree with their estimates. Our average-grain-size estimates for SAO are in better agreement with their results. \citet{carlson_distribution_2005} used log-normal distributions for the icy and acidic particle sizes and reported the radii, $r_m$, corresponding to the mean values of ln($r$). The effective scattering radius, $r_{eff}$, is the cross-section weighted radius \citep{hansen_light_1974} and for the log-normal case is $r_{eff} = r_m \textrm{exp}[(5/2)\sigma^2] \approx 3.0 r_m$, using the variance $\sigma^2 = 0.44$ found for Antarctic ice grains \citep{aitchison_lognormal_1957}. The corresponding SAO diameters they obtain for the $S1$, $S2$ and $S3$ cases are 37, 42, and 57 microns, respectively, and generally agree with the values we obtain ($S1: 42.21_{-4.61}^{+5.96}$ microns, $S2: 44.98_{-3.12}^{+3.23}$ microns, and $S3: 97.83_{-10.75}^{+11.87}$ microns). 

There can be many reasons for the discrepancy between our abundance estimates for SAO and those of \citet{carlson_distribution_2005}. Firstly, \citet{carlson_distribution_2005} used a model with only two components: sulfuric-acid-octahydrate and crystalline water ice. Amorphous water ice was not included due to the unavailability of its optical constants when they performed the work. Secondly, the reflectance model used in that work assumed spherical particles, while the radiative-transfer model we are using (see section \ref{sec:forward_model}) is better suited for irregular particles \citep{hapke_single-particle_2012} and is more likely to apply to airless planetary surfaces like Europa's. Our best direct evidence for irregular particles on airless bodies comes from microscopic studies of lunar regolith particles \citep[e.g][]{slyuta_physical_2014}. As for Europa, detailed surface photometry and Hapke modeling show that the average single particle phase function of Europan regolith particles are irregular in shape \citep{domingue_disk-resolved_1992, domingue_re-analysis_1997, hapke_single-particle_2012, verbiscer_photometric_2013}. Finally, to navigate around the problem of radiation-induced shift of the absorption band centers, which could have biased their results, \citet{carlson_distribution_2005} discarded the band-edge data near 1.4 and 1.9$\micron$. They also discarded data below 1.3 $\micron$ for all three spectra, to ensure consistent comparisons with other datasets used in their work that did not go below 1.3 $\micron$. We deal with the band-center shift problem by introducing a parameter that can perform that shift in wavelength, and hence are able to use all the data points in each spectrum.   

The strong detection of amorphous water ice in $S3$ (3.36$\sigma$ level) fits with our expectation. The prevalence of amorphous ice on the surface of Europa is well established, especially in regions of high particle irradiation \citep[e.g.][]{strazzulla_ion-beam-induced_1992} and/or colder temperatures \citep[e.g.][]{hansen_amorphous_2004}. $S3$ comes from the colder plain-terrains region in the high latitudes of the trailing hemisphere of Europa (see Figure \ref{fig:europa_map}). While particle irradiation is weaker at higher latitudes as compared to the equatorial latitudes \citep{nordheim_preservation_2018}, dominance of water at higher altitudes \citep{fanale_galileos_1999,hansen_amorphous_2004, carlson_distribution_2005} along with colder temperatures might enhance amorphous ice abundance. $S1$ shows weak evidence of amorphous ice (2.4$\sigma$ level) and comes from a region quite close to the equator. Perhaps the higher average temperatures in these regions -- that cause the annealing of amorphous ice to crystalline ice -- are dominating over the opposing effect of irradiation induced amorphization, resulting in only a minor presence of amorphous ice. $S2$ on the other hand, which also comes from equatorial latitudes, doesn't show any evidence for amorphous ice. This might be explained by the fact that $S2$ belongs more to the anti-Jovian hemisphere, where particle irradiation is weak beyond $\sim 20\degree$ latitudes, as compared to near the trailing hemisphere apex \citep{nordheim_preservation_2018}. Moreover, $S2$ also comes from a region containing numerous linea (image not shown here), implying high tidal stresses and perhaps heating from differential motions \citep{doggett_geologic_2009}, which would favour crystallization and work against irradiation induced amorphization. This potential form of enhanced heating could also explain why we see a very high SAO abundance of $\sim 99\%$ for $S2$, as heating of the surface on Europa has also been linked to the formation of sublimational lag deposits \citep{fagents_cryomagmatic_2000, carlson_distribution_2005}. Water has a higher vapor pressure as compared to sulphur allotropes and sulfuric acid, and even modest heating can saturate the surface with sulfurous material.

For crystalline ice, we obtain strong evidence for its presence in $S2$ (5.9$\sigma$) and $S3$ (4.76$\sigma$), consistent with the 1.65 $\micron$ absorption feature corresponding to crystalline ice being clearly visible in both the spectra (Figure \ref{fig:data}). Evidently, at the locations of $S2$ and $S3$, water-ice has not been amorphized to the depths being probed by reflectance spectroscopy in the 1-2.5 $\micron$ wavelength region. $S1$ on the other hand comes from the visibly darkest regions of the trailing hemisphere (Figure \ref{fig:europa_map}), quite close to the apex, where amorphization probably extends to the full sub-mm depths being probed by NIMS. This also relates well to the fact that only $S3$ shows strong evidence ($>3\sigma$) for both forms of water ice, probably because of its location in the high-latitude regions where the water-ice abundance is significantly higher than the equatorial latitudes in the trailing hemisphere. Hence, we are seeing both amorphous ice, formed due to irradiation and/or colder high-latitude temperatures, in the top-most layers of the regolith, and crystalline ice that is likely present just below it. Interestingly, although we retrieve very low abundances of crystalline ice for both $S2$ ($0.24_{-0.0007}^{+0.0012}\%$) and $S3$ ($0.36_{-0.0006}^{+0.0009}\%$), its effect on the spectra mainly comes from the very large average grain-size values of $382.44_{-70.66}^{+78.19}$ microns for $S2$ and  $919.14_{-96.88}^{+58.26}$ microns for $S3$. These large grain-sizes are consistent with the range of 10s-100s of microns that have been found in previous spectroscopic and polarimetric studies \citep[e.g.][]{dalton_europas_2012, ligier_vlt/sinfoni_2016, poch_polarimetry_2018}. It should be noted that it is more appropriate to think of these large grain-sizes as a collection of crystallites or grains, rather than one big crystal. While the initial production of hydrate would be uniform, sublimation of the more volatile H$_2$O would produce localized regions of water-rich and hydrate-rich material. The scale of these concentrations depends on photon path-lengths inside and outside the regolith, which is not easy to estimate. However, one can assume that the transport inside this regolith dominates, and hence the sizes of these regions can be parameterized as a pseudo-particle-size. 

The wavelength-shift parameter, $\Delta_{wav}$, shows a clear decreasing trend from $S1$ to $S3$. The shift is almost a channel-width: -0.0198$_{-0.0016}^{+0.0015}$ $\micron$ for $S1$, about half of that at -0.0098$_{-0.001}^{+0.001}$ $\micron$ for $S2$, and negligible at -0.0043$_{-0.0015}^{+0.0015}$ for $S3$ (the negative sign indicates a shift towards shorter wavelengths). As mentioned earlier, irradiation-induced wavelength shifts, especially of water absorption bands, are a well-studied phenomenon in the laboratory \citep[e.g.][]{nash_ios_1977, poston_spectral_2017, brunetto_characterizing_2020}. They are primarily caused by changes in the cubic structure of water ice \citep{baratta_31_1991}. If the shifts we infer are indeed correlated with the amount of irradiation experienced, the location corresponding to $S1$ has perhaps experienced the most irradiation among the three locations considered here. It should be noted that our simplified $\Delta_{wav}$ parameter shifts the whole spectrum, rather than just the regions near the water absorption bands, the latter being much more complicated to implement. However, as Figure \ref{fig:radiation_shift} shows, the shift mostly affects the fit near the water bands. The 1-2.5 $\micron$ wavelength range covered by our data has very little continuum and, although the fit near the continuum gets a little worse when the model is shifted, the overall impact is minimal. For data with a broader wavelength coverage, a more complicated shift treatment may be warranted.

Finally, for the filling factor parameter $\phi$, we retrieve a value of 0.15$_{-0.06}^{+0.07}$ for $S1$, 0.18$_{-0.06}^{+0.06}$ for $S2$ and 0.03$_{-0.01}^{+0.03}$ for $S3$. These values indicate regoliths of very high porosity, (1 - $\phi$), specifically 0.85$_{-0.07}^{+0.06}$ for $S1$, 0.81$_{-0.06}^{+0.06}$ for $S2$ and 0.97$_{-0.032}^{+0.016}$ for $S3$, respectively. For comparison to familiar terrestrial snow examples, fresh wind-blown snow has measured porosity values in the range of $83\%$ to $85\%$ \citep{Clifton_2007}, and the topmost layer of frigid, dry snow deposits can reach almost $91\%$ porosity \citep{Fu_2018}. Despite the complex correlation between $\phi$ and other parameters discussed earlier, particularly with the calibration parameter $c$, the retrieved distributions for $\phi$ are well constrained in all three cases (Figures \ref{fig:S1_results}-\ref{fig:S2_results}). Reported values for Europa's porosity in the literature are limited and have never employed spectroscopic analysis. Our relatively high values of porosity are generally consistent with measurements of the low thermal-interia of airless icy bodies as well as Europa's surface in particular \citep{2016A&A...588A.133F,2020Icar..33813500R, 2010Icar..210..763R}. \citet{nelson_laboratory_2018} show that laboratory polarization studies of airless solar system body surface analogues indicate that  the surface of Europa can have a porosity as high as 90\%. On the other hand, \citet{poch_polarimetry_2018} performed polarimetric studies of ice particles, with their analysis showing that Europa most likely has sintered grains, leading to a more compact or less porous regolith. However, they caution that, unlike their pure water-ice lab samples, Europa's surface contains a significant amount of dark non-icy materials, which can significantly affect the degree of polarization. Moreover, they did not directly study the effect of porosity on polarimetric phase curves. Fits of the most recent \citet{hapke_2012} model to combined Earth-based telescopic and spacecraft observations of Europa's visual photometric phase curve simultaneously constrain the porosity from the angular-width of the shadow-hiding opposition effect as well as from the way increased compaction states can amplify the reflectance overall.  The global-average fit reported in \citet{verbiscer_photometric_2013} constrains Europa’s porosity at around 0.61$\pm$0.08,\footnote{Note: \citet{verbiscer_photometric_2013} original estimate of Europa's porosity was reported as 0.53, but we have recalculated the numbers to account for the $K$ parameter’s effect on coherent-backscatter opposition surge, which is important at the low phase angles studied in their paper.} somewhat less porous than what we have obtained in our individual regions. However, our own results suggest that the porosity may vary significantly over different terrain-types and regions of Europa's surface.

Since reflectance spectroscopy at the \textit{Galileo} NIMS wavelengths is probing just the upper 100s of microns-millimeter of the regolith, it is reasonable to think that the surface we are seeing is highly porous or `fluffy' in a thin layer at the top, but with compactness varying with depth and perhaps increasing. Another explanation of the high porosities we retrieve might be poorly calibrated data. As Figure \ref{fig:S3_sensitivity} shows, our model can only match the low reflectance level of $S3$ with very low values of $\phi$ (and extremely high values of the crystalline ice grain-size). Finally, deficiencies in the Hapke model itself might be forcing the retrieved porosity to become extreme. \citet{helfenstein_testing_2011} tested Hapke’s photometric model’s ability to constrain porosity of lab samples through their phase curve data, finding preliminary evidence that the effects of porosity may be difficult to detect in high-albedo surfaces. However, laboratory work is required to asses how well the $K$ parameter constrains the porosity of an icy regolith using spectroscopy.


\section{Conclusions}\label{sec:conclusions}

\begin{figure}[htbp!]
\centering
\includegraphics[width=\textwidth]{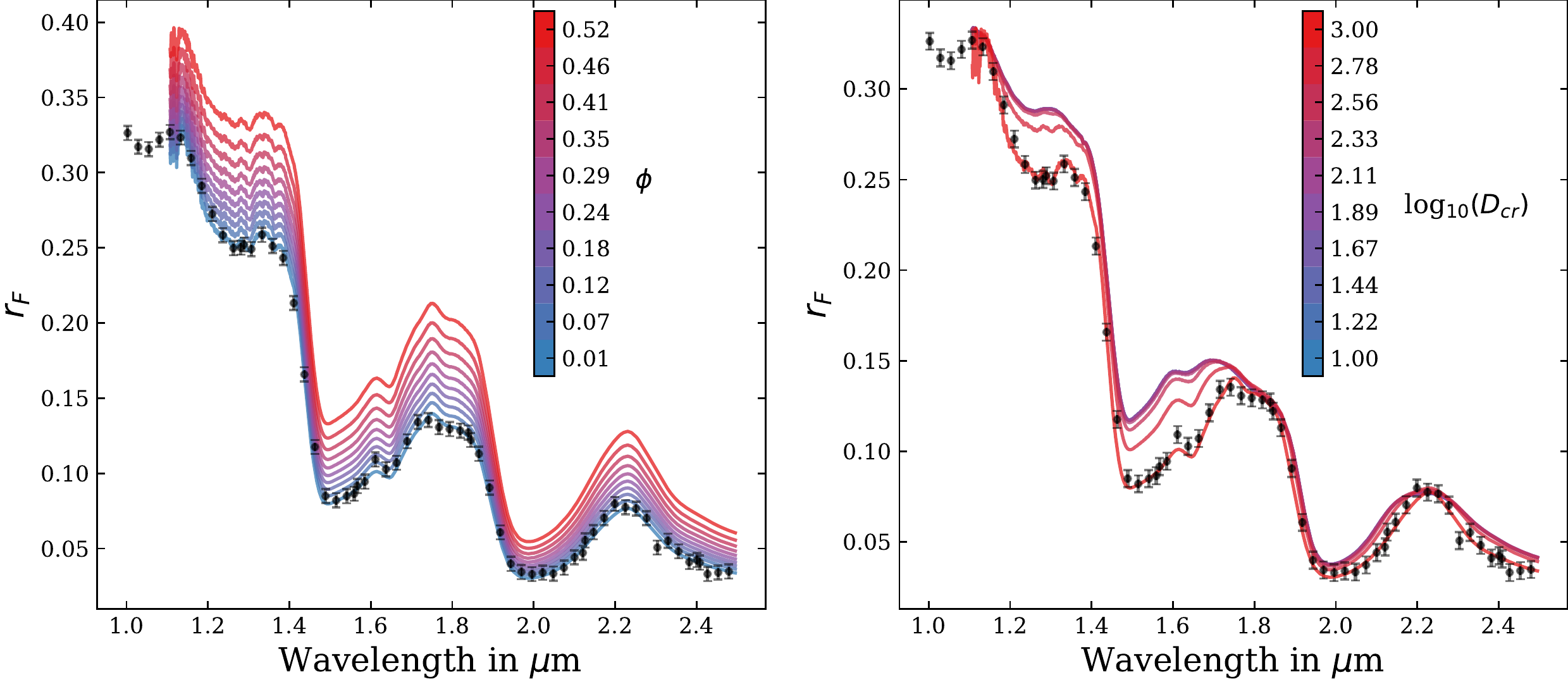}
\caption{Sensitivity of our best model for $S3$ to $\phi$ and $D_{cr}$. $\phi$ and $D_{cr}$ are varied in the left and right panels respectively, while all the other parameters are kept fixed to the retrieved median solution for $S3$ (Table \ref{tab:results_params}). The black points in both panels are the $S3$ data. Only for very small values of $\phi$ and very large values of $D_{cr}$, is the model able to match the low reflectance levels of the data. \label{fig:S3_sensitivity}}
\end{figure}

The highly non-linear radiative-transfer process through a planetary regolith calls for a radiative-transfer model, such as the Hapke model, to lie at the heart of spectroscopic analyses of planetary reflectance data. Here, we took the most up-to-date version of the Hapke model and make several quantitative improvements, relative to prior implementations, for Europan data analysis: (i) the simultaneous inclusion of amorphous and crystalline water ice, sulfuric-acid-octahydrate (SAO), CO$_2$, and SO$_2$; (ii) the use of physical parameters like regolith porosity; and (iii) consideration of a radiation-induced band-center shift and NIMS calibration-uncertainty. To thoroughly explore the complex multidimensional parameter space of the Hapke model, we employed a Bayesian inference framework. Most importantly, the Bayesian approach folds in both data uncertainties and model degeneracies into all parameter constraints, providing conservative estimates on key parameters of interested, such as species abundances. 
The key findings from the application of our framework to the three \emph{Galileo} NIMS spectra $S1$, $S2$ and $S3$ (Figures \ref{fig:data} \& \ref{fig:europa_map}) are as follows:

\begin{itemize}
    \item Sulfuric-acid-octahydrate or SAO is strongly detected (at $>30\sigma$ confidence level) and dominates the composition in all three spectra (Table \ref{tab:results_bmc}). This matches our expectations, as all three spectra come from the trailing hemisphere of Europa where sulfur-based hydrates are the dominant product of the surface radiolysis reactions. Among the three spectra, $S3$ has the lowest amount of SAO (Table \ref{tab:results_params}), which agrees with its physical location in the water-ice rich northern plain-terrains of the trailing hemisphere. 
    \item The presence of amorphous vs. crystalline water ice in the three spectra can possibly be understood through a balance of the opposing forces of heating (higher temperatures favor crystalline ice) and irradiation (higher irradiation levels favor amorphous ice). $S1$ shows weak evidence of amorphous ice (2.4$\sigma$) and no evidence of crystalline ice, which might indicate that the irradiation induced amorphization dominates over the higher than average temperatures near the trailing-hemishpere's apex. $S2$ shows no evidence for amorphous ice but a strong evidence for crystalline ice (5.9$\sigma$), consistent with the irradiation being weaker at $S2$'s location in the anti-Jovian hemisphere, despite the equatorial latitudes. Moreover, $S2$'s location has numerous linea, implying heating from differential motions, which favors crystallization. Finally, $S3$ shows strong evidence for both forms of ice, perhaps because it belongs to the cold northern plain-terrains where water ice is more abundant. The strong evidence for crystalline ice in $S2$ and $S3$ is also consistent with the presence of the 1.65 $\micron$ absorption feature seen clearly in the two spectra (Figure \ref{fig:data}). We note that this picture of the dominance of amorphous over crystalline ice and vice-versa, as a balance of particle irradiation and temperature, is may be a simplified picture of a complex set of factors that are yet to be well-understood.
    \item The grain-sizes retrieved for crystalline ice in $S2$ ($382.44_{-70.65}^{+78.19}$ microns) and $S3$ ($919.14_{-96.88}^{+58.26}$ microns) are rather large. We interpret these large grain-sizes as being representative of the scale of water ice concentrations formed due to the active sublimation of water-ice. Also, since our model retrieves these large grain-sizes, the corresponding abundances of crystalline ice for $S2$ ($0.24_{-0.0007}^{+0.0012}\%$) and $S3$ ($0.36_{-0.0006}^{+0.0009}\%$) are fairly low. These two parameters are usually highly anti-correlated, as evidenced in the 2D parameter distribution plot for $S2$ (Figure \ref{fig:S2_corner}).
    \item The wavelength-shift parameter, $\Delta_{wav}$, shows a clear decreasing trend from $S1$ to $S3$. The shift, towards shorter wavelengths, is almost a channel-width (0.0198$_{-0.0016}^{+0.0015}$ $\micron$ for $S1$, about half of that at 0.0098$_{-0.001}^{+0.001}$ $\micron$ for $S2$, and negligible at 0.0043$_{-0.0015}^{+0.0015}$ for $S3$). This fits well with the general expectation that $S1$ comes from the most irradiated, visibly dark, equatorial regions of Europa's trailing hemisphere, while $S3$ is from the northern latitudes where irradiation is much weaker.
    \item We retrieve high porosities for all three spectra. In particular, $S3$ requires a high porosity for the model to match the low reflectance levels of the data (Figure \ref{fig:S3_sensitivity}). While our porosities are consistent with previous studies that use polarimetric and photometric phase curves to constrain the porosity of Europa, more laboratory work is needed to test the ability of the $K$ parameter (porosity coefficient) in Hapke's model to constrain the porosity of an icy regolith using spectroscopic data.
\end{itemize}

Our work illustrates that the application of physically motivated reflectance models, in state-of-the-art statistical frameworks, can yield new insights from the archival \emph{Galileo} NIMS data of Europa. The technique of quantifying the evidence of a given species from the data, using Bayesian model comparison, holds promise for uncovering trace species abundances like organics on Europa, which is one of the primary goals for next generation missions like Europa Clipper \citep{blaney_mapping_2017} and JUICE \citep{grasset_jupiter_2013}. Crucially, like many studies before \citep[e.g.][]{dalton_spectroscopy_2010, dalton_europas_2012, ligier_vlt/sinfoni_2016,shirley_europas_2016, filacchione_serendipitous_2019, mishra_bayesian_2021}, our study highlights the need for laboratory measurements of cryogenic optical constants relevant to Europan conditions. The optical constants of irradiated samples are also important to accurately model the conditions on Europa. For example, we now know that irradiated sodium chloride exists on Europa due to laboratory work that demonstrated the development of color-centers in the salt when irradiated \citep{trumbo_sodium_2019}. As more optical constants of species relevant to Europa become available, we plan to incorporate them into our framework and revisit the \emph{Galileo} NIMS dataset, as well as NIR Europan data collected by other missions like \textit{Cassini} \citep{brown_observations_2003} and \textit{New Horizons} \citep{grundy_new_2007}, where plenty of new knowledge about the complex Europan surface composition might be awaiting.

\section{Software and third party data repository citations} \label{sec:software}

\textit{Repository:}  \url{sshade.eu} \citep{sshade}
\textit{Software:} NumPy \citep{harris2020array}, Jupyter \citep{Kluyver:2016aa}, Matplotlib \citep{Hunter:2007}, SciPy \citep{2020SciPy-NMeth}, pandas \citep{reback2020pandas}, dynesty \citep{speagle_dynesty_2020}, corner \citep{corner}, pandexo \citep{batalha_pandexo_2017}

\acknowledgements

\noindent IM is supported by the National Aeronautics and Space Administration (NASA) FINESST grant PLANET20-0103. JL is supported by Juno subcontract D99069MO to Cornell University from the Southwest Research Institute. PH thanks CCAPS and the Cornell University Visiting Scientist program for his participation. IM also thanks Samantha Trumbo for her kind assistance with plotting the geological maps. We also acknowledge that the data used in this work is available in NASA's Planetary Data System or PDS (Volume 1104 and files g1e003ci\_spc.cub (Radiance Factor) and g1e003ci\_loc.cub (Latitude \& Longitude)).


\clearpage
\bibliography{ms}{}

\begin{thebibliography}{}
\expandafter\ifx\csname natexlab\endcsname\relax\def\natexlab#1{#1}\fi
\providecommand{\url}[1]{\href{#1}{#1}}
\providecommand{\dodoi}[1]{doi:~\href{http://doi.org/#1}{\nolinkurl{#1}}}
\providecommand{\doeprint}[1]{\href{http://ascl.net/#1}{\nolinkurl{http://ascl.net/#1}}}
\providecommand{\doarXiv}[1]{\href{https://arxiv.org/abs/#1}{\nolinkurl{https://arxiv.org/abs/#1}}}

\bibitem[{Aitchison \& Brown(1957)}]{aitchison_lognormal_1957}
Aitchison, J., \& Brown, J. A.~C. 1957, The {Lognormal} {Distribution} with
  special reference to its uses in econometrics, 1st edn. (Cambridge: Cambridge
  University Press)

\bibitem[{Baratta {et~al.}(1991)Baratta, Leto, Spinella, Strazzulla, \&
  Foti}]{baratta_31_1991}
Baratta, G.~A., Leto, G., Spinella, F., Strazzulla, G., \& Foti, G. 1991,
  Astronomy and Astrophysics, 252, 421.
\newblock \url{http://adsabs.harvard.edu/abs/1991A%26A...252..421B}

\bibitem[{Batalha {et~al.}(2017)Batalha, Mandell, Pontoppidan, Stevenson,
  Lewis, Kalirai, Earl, Greene, Albert, \& Nielsen}]{batalha_pandexo_2017}
Batalha, N.~E., Mandell, A., Pontoppidan, K., {et~al.} 2017, Publications of
  the Astronomical Society of the Pacific, 129, 064501,
  \dodoi{10.1088/1538-3873/aa65b0}

\bibitem[{Belgacem {et~al.}(2020)Belgacem, Schmidt, \&
  Jonniaux}]{belgacem_regional_2020}
Belgacem, I., Schmidt, F., \& Jonniaux, G. 2020, Icarus, 338, 113525,
  \dodoi{10.1016/j.icarus.2019.113525}

\bibitem[{Benneke \& Seager(2012)}]{benneke_atmospheric_2012}
Benneke, B., \& Seager, S. 2012, The Astrophysical Journal, 753, 100,
  \dodoi{10.1088/0004-637X/753/2/100}

\bibitem[{Blaney {et~al.}(2017)Blaney, Clark, Dalton, Davies, Green, Hibbits,
  Langevin, Lunine, McCord, Paranicas, Murchie, Seelos, \&
  Soderblom}]{blaney_mapping_2017}
Blaney, D.~L., Clark, R.~N., Dalton, J.~B., {et~al.} 2017, Lunar and Planetary
  Science Conference, 2244.
\newblock
  \url{https://ui.adsabs.harvard.edu//#abs/2017LPI....48.2244B/abstract}

\bibitem[{Brown \& Hand(2013)}]{brown_salts_2013}
Brown, M.~E., \& Hand, K.~P. 2013, The Astronomical Journal, 145, 110,
  \dodoi{10.1088/0004-6256/145/4/110}

\bibitem[{Brown {et~al.}(2003)Brown, Baines, Bellucci, Bibring, Buratti,
  Capaccioni, Cerroni, Clark, Coradini, Cruikshank, Drossart, Formisano,
  Jaumann, Langevin, Matson, McCord, Mennella, Nelson, Nicholson, Sicardy,
  Sotin, Amici, Chamberlain, Filacchione, Hansen, Hibbitts, \&
  Showalter}]{brown_observations_2003}
Brown, R., Baines, K., Bellucci, G., {et~al.} 2003, Icarus, 164, 461,
  \dodoi{10.1016/S0019-1035(03)00134-9}

\bibitem[{Brunetto {et~al.}(2020)Brunetto, Lantz, Nakamura, Baklouti,
  Le~Pivert-Jolivet, Kobayashi, \& Borondics}]{brunetto_characterizing_2020}
Brunetto, R., Lantz, C., Nakamura, T., {et~al.} 2020, Icarus, 345, 113722,
  \dodoi{10.1016/j.icarus.2020.113722}

\bibitem[{Carlson {et~al.}(2005)Carlson, Anderson, Mehlman, \&
  Johnson}]{carlson_distribution_2005}
Carlson, R.~W., Anderson, M.~S., Mehlman, R., \& Johnson, R.~E. 2005, Icarus,
  177, 461, \dodoi{10.1016/j.icarus.2005.03.026}

\bibitem[{Carlson {et~al.}(2009)Carlson, Calvin, Dalton, Hansen, Hudson,
  Johnson, McCord, \& Moore}]{carlson_europas_2009}
Carlson, R.~W., Calvin, W.~M., Dalton, J.~B., {et~al.} 2009, Europa, 283.
\newblock \url{https://ui.adsabs.harvard.edu/abs/2009euro.book..283C/abstract}

\bibitem[{Carlson {et~al.}(1999{\natexlab{a}})Carlson, Johnson, \&
  Anderson}]{carlson_sulfuric_1999}
Carlson, R.~W., Johnson, R.~E., \& Anderson, M.~S. 1999{\natexlab{a}}, Science,
  286, 97, \dodoi{10.1126/science.286.5437.97}

\bibitem[{Carlson {et~al.}(1992)Carlson, Weissman, Smythe, \&
  Mahoney}]{carlson_near-infrared_1992}
Carlson, R.~W., Weissman, P.~R., Smythe, W.~D., \& Mahoney, J.~C. 1992, in The
  {Galileo} {Mission}, ed. C.~T. Russell (Dordrecht: Springer Netherlands),
  457--502, \dodoi{10.1007/978-94-011-2512-3_18}

\bibitem[{Carlson {et~al.}(1999{\natexlab{b}})Carlson, Anderson, Johnson,
  Smythe, Hendrix, Barth, Soderblom, Hansen, McCord, Dalton, Clark, Shirley,
  Ocampo, \& Matson}]{carlson_hydrogen_1999}
Carlson, R.~W., Anderson, M.~S., Johnson, R.~E., {et~al.} 1999{\natexlab{b}},
  Science, 283, 2062, \dodoi{10.1126/science.283.5410.2062}

\bibitem[{Carr {et~al.}(1998)Carr, Belton, Chapman, Davies, Geissler,
  Greenberg, McEwen, Tufts, Greeley, Sullivan, Head, Pappalardo, Klaasen,
  Johnson, Kaufman, Senske, Moore, Neukum, Schubert, Burns, Thomas, \&
  Veverka}]{carr_evidence_1998}
Carr, M.~H., Belton, M. J.~S., Chapman, C.~R., {et~al.} 1998, Nature, 391, 363,
  \dodoi{10.1038/34857}

\bibitem[{Chandrasekhar(1960)}]{chandrasekhar_radiative_1960}
Chandrasekhar, S. 1960, New York: Dover, 1960.
\newblock \url{http://adsabs.harvard.edu/abs/1960ratr.book.....C}

\bibitem[{Chyba \& Phillips(2001)}]{chyba_possible_2001}
Chyba, C.~F., \& Phillips, C.~B. 2001, Proceedings of the National Academy of
  Sciences, 98, 801, \dodoi{10.1073/pnas.98.3.801}

\bibitem[{Clark {et~al.}(2012)Clark, Cruikshank, Jaumann, Brown, Stephan,
  Dalle~Ore, Eric~Livo, Pearson, Curchin, Hoefen, Buratti, Filacchione, Baines,
  \& Nicholson}]{clark_surface_2012}
Clark, R.~N., Cruikshank, D.~P., Jaumann, R., {et~al.} 2012, Icarus, 218, 831,
  \dodoi{10.1016/j.icarus.2012.01.008}

\bibitem[{Clifton {et~al.}(2007)Clifton, Manes, Rüedi, Guala, \&
  Lehning}]{Clifton_2007}
Clifton, A., Manes, C., Rüedi, J.-D., Guala, M., \& Lehning, M. 2007,
  Boundary-Layer Meteorology, 126, 249, \dodoi{10.1007/s10546-007-9235-0}

\bibitem[{Dalton(2007)}]{dalton_linear_2007}
Dalton, J.~B. 2007, Geophysical Research Letters, 34,
  \dodoi{10.1029/2007GL031497}

\bibitem[{Dalton(2010)}]{dalton_spectroscopy_2010}
---. 2010, Space Science Reviews, 153, 219, \dodoi{10.1007/s11214-010-9658-7}

\bibitem[{Dalton {et~al.}(2012)Dalton, Shirley, \& Kamp}]{dalton_europas_2012}
Dalton, J.~B., Shirley, J.~H., \& Kamp, L.~W. 2012, Journal of Geophysical
  Research: Planets, 117, \dodoi{10.1029/2011JE003909}

\bibitem[{Doggett {et~al.}(2009)Doggett, Greeley, Figueredo, Tanaka, \&
  Dotson}]{doggett_geologic_2009}
Doggett, T., Greeley, R., Figueredo, P., Tanaka, K., \& Dotson, R. 2009, in
  Europa, ed. R.~T. Pappalardo, W.~B. McKinnon, \& K.~K. Khurana (University of
  Arizona Press), 137--160, \dodoi{10.2307/j.ctt1xp3wdw.12}

\bibitem[{Domingue \& Hapke(1992)}]{domingue_disk-resolved_1992}
Domingue, D., \& Hapke, B. 1992, Icarus, 99, 70,
  \dodoi{10.1016/0019-1035(92)90172-4}

\bibitem[{Domingue \& Verbiscer(1997)}]{domingue_re-analysis_1997}
Domingue, D., \& Verbiscer, A. 1997, Icarus, 128, 49,
  \dodoi{10.1006/icar.1997.5730}

\bibitem[{Dybwad(1971)}]{dybwad_radiation_1971}
Dybwad, J.~P. 1971, Journal of Geophysical Research (1896-1977), 76, 4023,
  \dodoi{https://doi.org/10.1029/JB076i017p04023}

\bibitem[{Fagents {et~al.}(2000)Fagents, Greeley, Sullivan, Pappalardo,
  Prockter, \& {Galileo SSI Team}}]{fagents_cryomagmatic_2000}
Fagents, S.~A., Greeley, R., Sullivan, R.~J., {et~al.} 2000, Icarus, 144, 54,
  \dodoi{10.1006/icar.1999.6254}

\bibitem[{Fanale {et~al.}(1999)Fanale, Granahan, McCord, Hansen, Hibbitts,
  Carlson, Matson, Ocampo, Kamp, Smythe, Leader, Mehlman, Greeley, Sullivan,
  Geissler, Barth, Hendrix, Clark, Helfenstein, Veverka, Belton, Becker,
  Becker, \& Galileo~NIMS}]{fanale_galileos_1999}
Fanale, F.~P., Granahan, J.~C., McCord, T.~B., {et~al.} 1999, Icarus, 139, 179,
  \dodoi{10.1006/icar.1999.6117}

\bibitem[{Fernando {et~al.}(2013)Fernando, Schmidt, Ceamanos, Pinet, Douté, \&
  Daydou}]{fernando_surface_2013}
Fernando, J., Schmidt, F., Ceamanos, X., {et~al.} 2013, Journal of Geophysical
  Research: Planets, 118, 534, \dodoi{10.1029/2012JE004194}

\bibitem[{Fernando {et~al.}(2016)Fernando, Schmidt, \&
  Douté}]{fernando_martian_2016}
Fernando, J., Schmidt, F., \& Douté, S. 2016, Planetary and Space Science,
  128, 30, \dodoi{10.1016/j.pss.2016.05.005}

\bibitem[{{Ferrari} \& {Lucas}(2016)}]{2016A&A...588A.133F}
{Ferrari}, C., \& {Lucas}, A. 2016, \aap, 588, A133,
  \dodoi{10.1051/0004-6361/201527625}

\bibitem[{Filacchione {et~al.}(2019)Filacchione, Adriani, Mura, Tosi, Lunine,
  Raponi, Ciarniello, Grassi, Piccioni, Moriconi, Altieri, Plainaki, Sindoni,
  Noschese, Cicchetti, Bolton, \& Brooks}]{filacchione_serendipitous_2019}
Filacchione, G., Adriani, A., Mura, A., {et~al.} 2019, Icarus, 328, 1,
  \dodoi{10.1016/j.icarus.2019.03.022}

\bibitem[{Foreman-Mackey(2016)}]{corner}
Foreman-Mackey, D. 2016, The Journal of Open Source Software, 1, 24,
  \dodoi{10.21105/joss.00024}

\bibitem[{Fu {et~al.}(2018)Fu, Peng, Li, Cui, Liu, Yan, \& Chen}]{Fu_2018}
Fu, Q., Peng, L., Li, T., {et~al.} 2018, Water Science and Technology: Water
  Supply, 19, ws2018096, \dodoi{10.2166/ws.2018.096}

\bibitem[{Grasset {et~al.}(2013)Grasset, Dougherty, Coustenis, Bunce, Erd,
  Titov, Blanc, Coates, Drossart, Fletcher, Hussmann, Jaumann, Krupp, Lebreton,
  Prieto-Ballesteros, Tortora, Tosi, \& Van~Hoolst}]{grasset_jupiter_2013}
Grasset, O., Dougherty, M.~K., Coustenis, A., {et~al.} 2013, Planetary and
  Space Science, 78, 1, \dodoi{10.1016/j.pss.2012.12.002}

\bibitem[{Greeley {et~al.}(2009)Greeley, Pappalardo, Prockter, Hendrix, Lock,
  \& Dotson}]{greeley_future_2009}
Greeley, R., Pappalardo, R.~T., Prockter, L.~M., {et~al.} 2009, in Europa, ed.
  R.~T. Pappalardo, W.~B. McKinnon, \& K.~K. Khurana (University of Arizona
  Press), 655--696, \dodoi{10.2307/j.ctt1xp3wdw.34}

\bibitem[{Grundy \& Schmitt(1998)}]{grundy_temperature-dependent_1998}
Grundy, W.~M., \& Schmitt, B. 1998, Journal of Geophysical Research: Planets,
  103, 25809, \dodoi{10.1029/98JE00738}

\bibitem[{Grundy {et~al.}(2007)Grundy, Buratti, Cheng, Emery, Lunsford,
  McKinnon, Moore, Newman, Olkin, Reuter, Schenk, Spencer, Stern, Throop, \&
  Weaver}]{grundy_new_2007}
Grundy, W.~M., Buratti, B.~J., Cheng, A.~F., {et~al.} 2007, Science, 318, 234,
  \dodoi{10.1126/science.1147623}

\bibitem[{Hand \& Carlson(2012)}]{hand2012carbon}
Hand, K., \& Carlson, R. 2012, Journal of Geophysical Research: Planets, 117

\bibitem[{Hand \& Brown(2013)}]{hand_keck_2013}
Hand, K.~P., \& Brown, M.~E. 2013, The Astrophysical Journal Letters, 766, L21,
  \dodoi{10.1088/2041-8205/766/2/L21}

\bibitem[{Hand {et~al.}(2007)Hand, Carlson, \& Chyba}]{hand_energy_2007}
Hand, K.~P., Carlson, R.~W., \& Chyba, C.~F. 2007, Astrobiology, 7, 1006,
  \dodoi{10.1089/ast.2007.0156}

\bibitem[{Hand {et~al.}(2006)Hand, Chyba, Carlson, \&
  Cooper}]{hand_clathrate_2006}
Hand, K.~P., Chyba, C.~F., Carlson, R.~W., \& Cooper, J.~F. 2006, Astrobiology,
  6, 463, \dodoi{10.1089/ast.2006.6.463}

\bibitem[{Hand \& German(2018)}]{hand_exploring_2018}
Hand, K.~P., \& German, C.~R. 2018, Nature Geoscience, 11, 2,
  \dodoi{10.1038/s41561-017-0045-9}

\bibitem[{Hansen \& McCord(2004)}]{hansen_amorphous_2004}
Hansen, G.~B., \& McCord, T.~B. 2004, Journal of Geophysical Research: Planets,
  109, \dodoi{10.1029/2003JE002149}

\bibitem[{Hansen \& McCord(2008)}]{hansen_widespread_2008}
---. 2008, Geophysical Research Letters, 35, \dodoi{10.1029/2007GL031748}

\bibitem[{Hansen \& Travis(1974)}]{hansen_light_1974}
Hansen, J.~E., \& Travis, L.~D. 1974, Space Science Reviews, 16, 527,
  \dodoi{10.1007/BF00168069}

\bibitem[{Hapke(2008)}]{hapke_bidirectional_2008}
Hapke, B. 2008, Icarus, 195, 918, \dodoi{10.1016/j.icarus.2008.01.003}

\bibitem[{Hapke(2012{\natexlab{a}})}]{hapke_2012}
---. 2012{\natexlab{a}}, Theory of Reflectance and Emittance Spectroscopy, 2nd
  edn. (Cambridge University Press), \dodoi{10.1017/CBO9781139025683}

\bibitem[{Hapke(2012{\natexlab{b}})}]{hapke_opposition_2012}
---. 2012{\natexlab{b}}, in Theory of {Reflectance} and {Emittance}
  {Spectroscopy}, 2nd edn. (Cambridge: Cambridge University Press), 221--262,
  \dodoi{10.1017/CBO9781139025683.009}

\bibitem[{Hapke(2012{\natexlab{c}})}]{hapke_single-particle_2012}
---. 2012{\natexlab{c}}, in Theory of {Reflectance} and {Emittance}
  {Spectroscopy}, 2nd edn. (Cambridge: Cambridge University Press), 100--144,
  \dodoi{10.1017/CBO9781139025683.006}

\bibitem[{Hapke(2021)}]{Hapke_2021_opposition}
---. 2021, Icarus, 354, 114105, \dodoi{10.1016/j.icarus.2020.114105}

\bibitem[{Harris {et~al.}(2020)Harris, Millman, van~der Walt, Gommers,
  Virtanen, Cournapeau, Wieser, Taylor, Berg, Smith, Kern, Picus, Hoyer, van
  Kerkwijk, Brett, Haldane, del R{'{\i}}o, Wiebe, Peterson,
  G{'{e}}rard-Marchant, Sheppard, Reddy, Weckesser, Abbasi, Gohlke, \&
  Oliphant}]{harris2020array}
Harris, C.~R., Millman, K.~J., van~der Walt, S.~J., {et~al.} 2020, Nature, 585,
  357, \dodoi{10.1038/s41586-020-2649-2}

\bibitem[{Helfenstein \& Shepard(2011)}]{helfenstein_testing_2011}
Helfenstein, P., \& Shepard, M.~K. 2011, Icarus, 215, 83,
  \dodoi{10.1016/j.icarus.2011.07.002}

\bibitem[{Hendrix {et~al.}(2008)Hendrix, Carlson, \&
  Johnson}]{hendrix_europas_2008}
Hendrix, A., Carlson, R., \& Johnson, R. 2008, 40, 59.07.
\newblock \url{http://adsabs.harvard.edu/abs/2008DPS....40.5907H}

\bibitem[{Henyey \& Greenstein(1941)}]{henyey_diffuse_1941}
Henyey, L.~G., \& Greenstein, J.~L. 1941, The Astrophysical Journal, 93, 70,
  \dodoi{10.1086/144246}

\bibitem[{Higson {et~al.}(2019)Higson, Handley, Hobson, \&
  Lasenby}]{higson_dynamic_2019}
Higson, E., Handley, W., Hobson, M., \& Lasenby, A. 2019, Statistics and
  Computing, 29, 891, \dodoi{10.1007/s11222-018-9844-0}

\bibitem[{Hunter(2007)}]{Hunter:2007}
Hunter, J.~D. 2007, Computing in Science \& Engineering, 9, 90,
  \dodoi{10.1109/MCSE.2007.55}

\bibitem[{{Jeffreys}(1939)}]{jeffreys_1939}
{Jeffreys}, H. 1939, {The Theory of Probability}

\bibitem[{Kattenhorn \& Prockter(2014)}]{kattenhorn_evidence_2014}
Kattenhorn, S.~A., \& Prockter, L.~M. 2014, Nature Geoscience, 7, 762,
  \dodoi{10.1038/ngeo2245}

\bibitem[{Kluyver {et~al.}(2016)Kluyver, Ragan-Kelley, P{\'e}rez, Granger,
  Bussonnier, Frederic, Kelley, Hamrick, Grout, Corlay, Ivanov, Avila, Abdalla,
  \& Willing}]{Kluyver:2016aa}
Kluyver, T., Ragan-Kelley, B., P{\'e}rez, F., {et~al.} 2016, in Positioning and
  Power in Academic Publishing: Players, Agents and Agendas, ed. F.~Loizides \&
  B.~Schmidt, IOS Press, 87 -- 90

\bibitem[{Lapotre {et~al.}(2017)Lapotre, Ehlmann, \&
  Minson}]{lapotre_probabilistic_2017}
Lapotre, M. G.~A., Ehlmann, B.~L., \& Minson, S.~E. 2017, Journal of
  Geophysical Research: Planets, 122, 983, \dodoi{10.1002/2016JE005248}

\bibitem[{Ligier {et~al.}(2016)Ligier, Poulet, Carter, Brunetto, \&
  Gourgeot}]{ligier_vlt/sinfoni_2016}
Ligier, N., Poulet, F., Carter, J., Brunetto, R., \& Gourgeot, F. 2016, The
  Astronomical Journal, 151, 163, \dodoi{10.3847/0004-6256/151/6/163}

\bibitem[{Mastrapa {et~al.}(2009)Mastrapa, Sandford, Roush, Cruikshank, \&
  Dalle~Ore}]{mastrapa_optical_2009}
Mastrapa, R.~M., Sandford, S.~A., Roush, T.~L., Cruikshank, D.~P., \&
  Dalle~Ore, C.~M. 2009, The Astrophysical Journal, 701, 1347,
  \dodoi{10.1088/0004-637X/701/2/1347}

\bibitem[{McCord {et~al.}(1998)McCord, Hansen, Fanale, Carlson, Matson,
  Johnson, Smythe, Crowley, Martin, Ocampo, Hibbitts, Granahan, \&
  Team}]{mccord_salts_1998}
McCord, T.~B., Hansen, G.~B., Fanale, F.~P., {et~al.} 1998, Science, 280, 1242,
  \dodoi{10.1126/science.280.5367.1242}

\bibitem[{McCord {et~al.}(1999)McCord, Hansen, Matson, Johnson, Crowley,
  Fanale, Carlson, Smythe, Martin, Hibbitts, Granahan, \&
  Ocampo}]{mccord_hydrated_1999}
McCord, T.~B., Hansen, G.~B., Matson, D.~L., {et~al.} 1999, Journal of
  Geophysical Research: Planets, 104, 11827, \dodoi{10.1029/1999JE900005}

\bibitem[{Mishra {et~al.}(2021)Mishra, Lewis, Lunine, Helfenstein, MacDonald,
  Filacchione, \& Ciarniello}]{mishra_bayesian_2021}
Mishra, I., Lewis, N., Lunine, J., {et~al.} 2021, Icarus, 357, 114215,
  \dodoi{10.1016/j.icarus.2020.114215}

\bibitem[{Moore {et~al.}(2009)Moore, Black, Buratti, Phillips, Spencer,
  Sullivan, \& Dotson}]{moore_surface_2009}
Moore, J.~M., Black, G., Buratti, B., {et~al.} 2009, in Europa, ed. R.~T.
  Pappalardo, W.~B. McKinnon, \& K.~K. Khurana (University of Arizona Press),
  329--350, \dodoi{10.2307/j.ctt1xp3wdw.19}

\bibitem[{Nash \& Fanale(1977)}]{nash_ios_1977}
Nash, D.~B., \& Fanale, F.~P. 1977, Icarus, 31, 40,
  \dodoi{10.1016/0019-1035(77)90070-7}

\bibitem[{Nelson {et~al.}(2018)Nelson, Boryta, Hapke, Manatt, Shkuratov,
  Psarev, Vandervoort, Kroner, Nebedum, Vides, \&
  Quiñones}]{nelson_laboratory_2018}
Nelson, R.~M., Boryta, M.~D., Hapke, B.~W., {et~al.} 2018, Icarus, 302, 483,
  \dodoi{10.1016/j.icarus.2017.11.021}

\bibitem[{Nordheim {et~al.}(2018)Nordheim, Hand, \&
  Paranicas}]{nordheim_preservation_2018}
Nordheim, T.~A., Hand, K.~P., \& Paranicas, C. 2018, Nature Astronomy, 2, 673,
  \dodoi{10.1038/s41550-018-0499-8}

\bibitem[{pandas~development team(2020)}]{reback2020pandas}
pandas~development team, T. 2020, pandas-dev/pandas: Pandas, latest,  Zenodo,
  \dodoi{10.5281/zenodo.3509134}

\bibitem[{Poch {et~al.}(2018)Poch, Cerubini, Pommerol, Jost, \&
  Thomas}]{poch_polarimetry_2018}
Poch, O., Cerubini, R., Pommerol, A., Jost, B., \& Thomas, N. 2018, Journal of
  Geophysical Research (Planets), 123, 2564, \dodoi{10.1029/2018JE005753}

\bibitem[{Poston {et~al.}(2017)Poston, Carlson, \& Hand}]{poston_spectral_2017}
Poston, M.~J., Carlson, R.~W., \& Hand, K.~P. 2017, Journal of Geophysical
  Research: Planets, 122, 2644, \dodoi{10.1002/2017JE005429}

\bibitem[{Quirico {et~al.}(1999)Quirico, Douté, Schmitt, de~Bergh, Cruikshank,
  Owen, Geballe, \& Roush}]{quirico_composition_1999}
Quirico, E., Douté, S., Schmitt, B., {et~al.} 1999, Icarus, 139, 159,
  \dodoi{10.1006/icar.1999.6111}

\bibitem[{Quirico \& Schmitt(1997)}]{quirico_near-infrared_1997}
Quirico, E., \& Schmitt, B. 1997, Icarus, 127, 354,
  \dodoi{10.1006/icar.1996.5663}

\bibitem[{Rampe {et~al.}(2018)Rampe, Lapotre, Bristow, Arvidson, Morris,
  Achilles, Weitz, Blake, Ming, Morrison, Vaniman, Chipera, Downs, Grotzinger,
  Hazen, Peretyazhko, Sutter, Tu, Yen, Horgan, Castle, Craig, Des~Marais,
  Farmer, Gellert, McAdam, Morookian, Sarrazin, \& Treiman}]{rampe_sand_2018}
Rampe, E.~B., Lapotre, M. G.~A., Bristow, T.~F., {et~al.} 2018, Geophysical
  Research Letters, 45, 9488, \dodoi{10.1029/2018GL079073}

\bibitem[{{Rathbun} {et~al.}(2010){Rathbun}, {Rodriguez}, \&
  {Spencer}}]{2010Icar..210..763R}
{Rathbun}, J.~A., {Rodriguez}, N.~J., \& {Spencer}, J.~R. 2010, \icarus, 210,
  763, \dodoi{10.1016/j.icarus.2010.07.017}

\bibitem[{{Rathbun} \& {Spencer}(2020)}]{2020Icar..33813500R}
{Rathbun}, J.~A., \& {Spencer}, J.~R. 2020, \icarus, 338, 113500,
  \dodoi{10.1016/j.icarus.2019.113500}

\bibitem[{Schmidt \& Fernando(2015)}]{schmidt_realistic_2015}
Schmidt, F., \& Fernando, J. 2015, Icarus, 260, 73,
  \dodoi{10.1016/j.icarus.2015.07.002}

\bibitem[{Schmitt {et~al.}(2018)Schmitt, Bollard, Albert, Garenne, Gorbacheva,
  Bonal, \& Volcke}]{sshade}
Schmitt, B., Bollard, P., Albert, D., {et~al.} 2018, Solid Spectroscopy Hosting
  Architecture of Databases and Expertise, \dodoi{10.26302/SSHADE}

\bibitem[{Schmitt {et~al.}(1994)Schmitt, de~Bergh, Lellouch, Maillard, Barbe,
  \& Doute}]{schmitt_identification_1994}
Schmitt, B., de~Bergh, C., Lellouch, E., {et~al.} 1994, Icarus, 111, 79,
  \dodoi{10.1006/icar.1994.1135}

\bibitem[{Schmitt {et~al.}(1998)Schmitt, Quirico, Trotta, \&
  Grundy}]{schmitt_optical_1998}
Schmitt, B., Quirico, E., Trotta, F., \& Grundy, W.~M. 1998, in Solar {System}
  {Ices}: {Based} on {Reviews} {Presented} at the {International} {Symposium}
  “{Solar} {System} {Ices}” held in {Toulouse}, {France}, on {March}
  27–30, 1995, ed. B.~Schmitt, C.~De~Bergh, \& M.~Festou, Astrophysics and
  {Space} {Science} {Library} (Dordrecht: Springer Netherlands), 199--240,
  \dodoi{10.1007/978-94-011-5252-5_9}

\bibitem[{Shirley {et~al.}(2016)Shirley, Jamieson, \&
  Dalton}]{shirley_europas_2016}
Shirley, J.~H., Jamieson, C.~S., \& Dalton, J.~B. 2016, Earth and Space
  Science, 3, 326, \dodoi{10.1002/2015EA000149}

\bibitem[{Skilling(2006)}]{skilling_nested_2006}
Skilling, J. 2006, Bayesian Analysis, 1, 833, \dodoi{10.1214/06-BA127}

\bibitem[{Slyuta(2014)}]{slyuta_physical_2014}
Slyuta, E.~N. 2014, Solar System Research, 48, 330,
  \dodoi{10.1134/S0038094614050050}

\bibitem[{Speagle(2020)}]{speagle_dynesty_2020}
Speagle, J.~S. 2020, Monthly Notices of the Royal Astronomical Society, 493,
  3132, \dodoi{10.1093/mnras/staa278}

\bibitem[{Spencer {et~al.}(1999)Spencer, Tamppari, Martin, \&
  Travis}]{spencer_temperatures_1999}
Spencer, J.~R., Tamppari, L.~K., Martin, T.~Z., \& Travis, L.~D. 1999, Science,
  284, 1514, \dodoi{10.1126/science.284.5419.1514}

\bibitem[{Strazzulla(2011)}]{fischer_spatially_2015}
Strazzulla, G. 2011, Nuclear Instruments and Methods in Physics Research B,
  150, 842, \dodoi{10.1088/0004-6256/150/5/164}

\bibitem[{Strazzulla {et~al.}(1992)Strazzulla, Baratta, Leto, \&
  Foti}]{strazzulla_ion-beam-induced_1992}
Strazzulla, G., Baratta, G.~A., Leto, G., \& Foti, G. 1992, EPL (Europhysics
  Letters), 18, 517, \dodoi{10.1209/0295-5075/18/6/008}

\bibitem[{Trotta(2017)}]{trotta_bayesian_2017}
Trotta, R. 2017, arXiv:1701.01467 [astro-ph, stat].
\newblock \url{http://arxiv.org/abs/1701.01467}

\bibitem[{Trumbo {et~al.}(2019{\natexlab{a}})Trumbo, Brown, \&
  Hand}]{trumbo_sodium_2019}
Trumbo, S.~K., Brown, M.~E., \& Hand, K.~P. 2019{\natexlab{a}}, Science
  Advances, 5, eaaw7123, \dodoi{10.1126/sciadv.aaw7123}

\bibitem[{Trumbo {et~al.}(2019{\natexlab{b}})Trumbo, Brown, \&
  Hand}]{trumbo_h2o2_2019}
---. 2019{\natexlab{b}}, The Astronomical Journal, 158, 127,
  \dodoi{10.3847/1538-3881/ab380c}

\bibitem[{Verbiscer {et~al.}(2013)Verbiscer, Helfenstein, \&
  Buratti}]{verbiscer_photometric_2013}
Verbiscer, A.~J., Helfenstein, P., \& Buratti, B.~J. 2013, in The {Science} of
  {Solar} {System} {Ices}, ed. M.~S. Gudipati \& J.~Castillo-Rogez,
  Astrophysics and {Space} {Science} {Library} (New York, NY: Springer),
  47--72, \dodoi{10.1007/978-1-4614-3076-6_2}

\bibitem[{Virtanen {et~al.}(2020)Virtanen, Gommers, Oliphant, Haberland, Reddy,
  Cournapeau, Burovski, Peterson, Weckesser, Bright, {van der Walt}, Brett,
  Wilson, Millman, Mayorov, Nelson, Jones, Kern, Larson, Carey, Polat, Feng,
  Moore, {VanderPlas}, Laxalde, Perktold, Cimrman, Henriksen, Quintero, Harris,
  Archibald, Ribeiro, Pedregosa, {van Mulbregt}, \& {SciPy 1.0
  Contributors}}]{2020SciPy-NMeth}
Virtanen, P., Gommers, R., Oliphant, T.~E., {et~al.} 2020, Nature Methods, 17,
  261, \dodoi{10.1038/s41592-019-0686-2}

\end{thebibliography}
\bibliographystyle{aasjournal}



\end{document}